\documentclass[english,prd,twocolumn,superscriptaddress,floatfix,nofootinbib,preprintnumbers,eqsecnum]{revtex4-1}
\usepackage{graphicx}
\usepackage{amsmath,amsthm,amssymb}
\usepackage{bm}

\usepackage{amsfonts}
\usepackage{dcolumn}
\usepackage{hyperref}

\def\be{\begin{equation}}
\def\ee{\end{equation}}
\def\ba{\begin{eqnarray}}
\def\ea{\end{eqnarray}}
\def\bs{\begin{subequations}}
\def\es{\end{subequations}}

\renewcommand{\S}{{\text{\tiny ${\cal S}$}}}

\newcommand{\W}{{\text{\tiny ${\cal W}$}}}

\usepackage{color}

\newcommand{\s}{{\rm s}}
\newcommand{\tm}{{\rm t}}

\makeatother

\usepackage{babel}
\makeatother

\begin{document}

\title{Possibility of realizing weak gravity 
in redshift space distortion measurements}

\author{Shinji Tsujikawa}

\affiliation{Department of Physics, Faculty of Science, 
Tokyo University of Science,
1-3, Kagurazaka, Shinjuku-ku, Tokyo 162-8601, Japan}

\begin{abstract}

We study the possibility of realizing a growth rate of matter density 
perturbations lower than that in General Relativity.
Using the approach of the effective field theory of modified 
gravity encompassing theories beyond Horndeski, 
we derive the effective gravitational coupling $G_{\rm eff}$ 
and the gravitational slip parameter $\eta$ for perturbations 
deep inside the Hubble radius.
In Horndeski theories we derive a necessary condition 
for achieving weak gravity associated with tensor 
perturbations, but this is not a sufficient condition due to the presence 
of a scalar-matter interaction that always enhances $G_{\rm eff}$.  
Beyond the Horndeski domain it is possible to 
realize $G_{\rm eff}$ smaller than Newton's gravitational constant 
$G$, while the scalar and tensor perturbations satisfy no-ghost 
and stability conditions. 
We present a concrete dark energy 
scenario with varying $c_{\rm t}$ and numerically 
study the evolution of perturbations to confront the model 
with the observations of redshift-space 
distortions and weak lensing.

\end{abstract}

\date{\today}


\maketitle

\section{Introduction}

The observations of redshift-space distortions (RSD) and 
weak lensing \cite{2dF,6dF,Wiggle,SDSS,BOSS,Torre,JPAS,fastsound}, 
combined with Cosmic Microwave Background (CMB) 
measurements \cite{Planck1}, offer the possibility of 
testing General Relativity (GR) on cosmological scales. 
In particular, the observational evidence of late-time cosmic 
acceleration \cite{Riess} may be related to some modification of gravity 
at large distances. The dark energy equation of state 
$w_{\rm DE}<-1$, which is allowed from the joint 
analysis of CMB and supernovae Ia (SN Ia) data \cite{Planck1}, 
can be realized in modified gravitational theories 
without ghosts and instabilities \cite{moreview}.

If we modify gravity from GR, an extra scalar 
degree of freedom usually emerges due to the breaking of 
gauge symmetries of GR \cite{frreview}. 
This scalar field mediates an extra 
gravitational force with a matter sector. 
In $f(R)$ gravity, for example, the effective 
gravitational coupling $G_{\rm eff}$ between the gravitational 
scalar and matter is $4/3$ times as large as 
Newton's gravitational constant $G$ in the regime where 
the scalar mass $M$ is much smaller than the 
physical momentum $k/a$ of interest \cite{frper}.

The recent observations of RSD \cite{Beu,Ledo,Eriksen} and 
cluster counts \cite{Vik} have measured the lower growth 
rate of matter density perturbations $\delta_m$ than that 
predicted by the $\Lambda$-Cold-Dark-Matter ($\Lambda$CDM) 
model. In fact, the Planck CMB measurements \cite{Planck1,Planckmo} 
are in tension with the RSD data and the Hubble expansion data from SN Ia. 
One possibility for reconciling this discrepancy is to incorporate 
massive neutrinos \cite{Moss}, but this increases the tension between 
the CMB and the Hubble expansion measurements \cite{Mena,Planck2}.

Another possibility for realizing a lower cosmic growth rate 
is interacting models of vacuum energy and 
dark matter \cite{Freese,Lima,Pavon,Berman,Wands} 
(see also Ref.~\cite{Bertolami}).
If there is an energy transfer from dark matter to dark energy, 
it is possible to reduce the tension between the CMB and 
RSD measurements \cite{Salva}. 
Most of these interacting models are 
based on a phenomenological approach, in that the equations of 
motion do not follow from a concrete Lagrangian.
In this case, even if a lower growth rate consistent with 
observations is realized, it is not generally clear whether 
theoretically consistent conditions such as the 
absence of ghosts are satisfied or not.

There exists a modified gravitational scenario--dubbed the 
Dvali-Gabadadze-Porrati (DGP) braneworld model \cite{DGP}-- 
which possesses an explicit Lagrangian 
in the five-dimensional bulk space-time. 
In the branch where the late-time cosmic 
acceleration occurs, it is known that the effective 
gravitational coupling $G_{\rm eff}$ 
is smaller than $G$ on scales relevant to large-scale 
structures \cite{KM}.
However, ghosts are present in this accelerating branch.
Thus, in the DGP model, the lower cosmic growth rate 
is related to the appearance of ghosts \cite{dgpghost}.

Now, a question arises. 
Are there some modified gravity models 
with concrete Lagrangians realizing weak gravity 
($G_{\rm eff}$ smaller than $G$) on cosmological scales, 
while avoiding the ghosts 
and instabilities associated with the propagation speeds of 
scalar and tensor perturbations? 
In order to address this problem, we focus on a very general 
class of scalar-tensor theories dubbed 
Gleyzes-Langlois-Piazza-Vernizzi (GLPV) theories \cite{GLPV}.
This also accommodates Horndeski theories \cite{Horndeski}--the 
most general scalar-tensor theories 
with second-order equations of motion in a general space-time. 

The action of GLPV theories has been derived
in such a way that the Horndeski action written in the framework 
of the Arnowitt-Deser-Misner (ADM) formalism \cite{Gleyzes13} 
does not obey two additional conditions. 
While this can generate derivatives higher than second order 
in a generic space-time, there is no extra propagating degree of 
freedom on a flat 
Friedmann-Lema\^{i}tre-Robertson-Walker (FLRW) 
background according to Hamiltonian analysis 
in terms of cosmological perturbations \cite{Hami,Hami2}. 
This conclusion also holds for odd-mode perturbations
on a spherically symmetric background \cite{KGT}.

In GLPV theories the tensor propagation speed squared $c_{\tm}^2$ 
can deviate from 1 even during the radiation- and early matter-dominated 
epochs, while in Horndeski theories this is limited 
by the two extra conditions mentioned above \cite{Gleyzes13,DeFelice}. 
Even for a simple canonical scalar field $\phi$ with a potential, 
the deviation of $c_{\tm}^2$ from 1 can give rise to an 
interesting observational signature such as a large difference 
between two gravitational potentials $\Psi$ and $\Phi$ \cite{DKT}. 
For constant $c_{\rm t}^2$ models one has
$G_{\rm eff}<G$ in the superluminal regime ($c_{\rm t}^2>1$), 
but $G_{\rm eff}$ needs to be very close to $G$ due to
the fact that the scalar propagation speed 
squared $c_{\rm s}^2$ becomes negative as $c_{\rm t}^2$ 
is away from 1.
In this paper we show that it is possible to realize weak gravity 
in varying $c_{\rm t}^2$ models, 
while satisfying the no-ghost and stability conditions associated 
with scalar and tensor perturbations.

In Sec.~\ref{modelsec} we begin with a brief review of 
the effective field theory of modified gravity \cite{Weinberg,Park,Bloomfield,Battye,Mueller,Gubi,Piazza2,Silve,Gergely,Tsuji15,Gao,Kasecosmo,Kase,GLP} in which GLPV theories are encompassed as a specific case.
In the presence of a matter component, we present the background and 
linear perturbation equations of motion in the unitary gauge in Sec.~\ref{persec}. 

In Sec.~\ref{subsec} we derive the effective gravitational 
coupling $G_{\rm eff}/G$ and the gravitational slip 
parameter $\eta=-\Phi/\Psi$ by employing a subhorizon approximation
for the perturbations relevant to large-scale structures.
In GLPV theories time derivatives of metric perturbations are left even under 
this approximation, so the usual quasistatic approximation is trustable 
only when these time derivatives are suppressed relative to other terms. 
In this case, we implement such time derivatives as corrections to 
leading-order terms.

In Sec.~\ref{weaksec} we discuss the possibility of realizing weak gravity 
by expressing $G_{\rm eff}/G$ and $\eta$ in terms of quantities 
associated with the no-ghost and stability conditions of 
scalar and tensor perturbations.
In Horndeski theories we derive a necessary condition for 
realizing $G_{\rm eff}<G$, which is related with 
quantities appearing in the second-order action of tensor 
perturbations. However, this is not a sufficient condition 
due to an extra scalar interaction with matter which 
always enhances $G_{\rm eff}$.

In GLPV theories the value of $c_{\tm}^2$ is not restricted to be 
close to 1 even in the early cosmological epoch.
In Sec.~\ref{GLPVsec} we propose a simple model 
with a time-varying $c_{\tm}^2$ in which 
the realization of weak gravity is possible 
without ghosts and Laplacian instabilities.
In Sec.~\ref{obsersec} we numerically solve the full perturbation 
equations of motion for the decreasing $c_{\tm}^2$ model 
with $0<c_{\tm}^2<1$ and show that the growth rate of matter 
perturbations associated with RSD measurements can be lower  
than that predicted by the $\Lambda$CDM model.

\section{Modified gravitational theories}
\label{modelsec}

The most general scalar-tensor theories with second-order equations of 
motion are known as Horndeski theories \cite{Horndeski}.
The four-dimensional action of Horndeski theories is given by 
$S=\int d^4 x \sqrt{-g}\,L$ with the 
Lagrangian \cite{DGSZ,Koba,Char}
\ba
L &=& G_2(\phi,X)+G_{3}(\phi,X)\square\phi \nonumber \\
&& +G_{4}(\phi,X)\, R-2G_{4,X}(\phi,X)\left[ (\square \phi)^{2}
-\phi^{;\mu \nu }\phi _{;\mu \nu} \right] \nonumber \\
& &
+G_{5}(\phi,X)G_{\mu \nu }\phi ^{;\mu \nu}
+\frac{1}{3}G_{5,X}(\phi,X)
[ (\square \phi )^{3} \nonumber \\
& &-3(\square \phi )\,\phi _{;\mu \nu }\phi ^{;\mu
\nu }+2\phi _{;\mu \nu }\phi ^{;\mu \sigma }{\phi ^{;\nu}}_{;\sigma}]\,,
\label{Lho}
\ea
where $g$ is a determinant of the four-dimensional metric 
$g_{\mu \nu}$, $G_{2,3,4,5}$ are functions in terms 
of a scalar field $\phi$ and its kinetic energy 
$X=g^{\mu \nu} \phi_{;\mu} \phi_{;\nu}$, 
$R$ and $G_{\mu\nu}$ are the four-dimensional Ricci scalar and 
the Einstein tensor respectively, and a semicolon 
represents a covariant derivative with
$\square \phi \equiv (g^{\mu \nu} \phi_{;\nu})_{;\mu}$.

GLPV theories \cite{GLPV} correspond to the generalization 
of Horndeski theories derived by reformulating
the Lagrangian (\ref{Lho}) 
in terms of the 3+1 ADM decomposition of space-time \cite{ADM} 
with the foliation of constant-time hypersurfaces $\Sigma_t$. 
The ADM formalism is based upon the line element 
$ds^2=g_{\mu \nu}dx^{\mu}dx^{\nu}= 
-N^{2}dt^{2}+h_{ij}(dx^{i}+N^{i}dt)(dx^{j}+N^{j}dt)$, 
where $N$ is the lapse, $N^i$ is the shift, and 
$h_{ij}$ is the three-dimensional spatial metric.

The extrinsic curvature and the intrinsic curvature 
are defined, respectively, by  
$K_{\mu \nu}=h^{\lambda}_{\mu} n_{\nu;\lambda}$ and 
${\cal R}_{\mu \nu}={}^{(3)}R_{\mu \nu}$, where
$n_{\mu}=(-N,0,0,0)$ is a normal vector orthogonal 
to $\Sigma_t$ and ${}^{(3)}R_{\mu \nu}$ is the 
three-dimensional Ricci tensor on $\Sigma_t$.
In the following we shall focus on a flat FLRW background 
described by the line element $ds^2=-dt^2+a^2(t)\delta_{ij}dx^idx^j$, 
where $a(t)$ is the scale factor.
In the unitary gauge the scalar field $\phi$ 
depends on the time $t$ alone and 
hence $X=-N^{-2}\dot{\phi}^2$, 
where a dot represents a derivative with respect to $t$.
For this gauge choice, the action of GLPV theories 
can be written as 
\ba
S &=&
\int d^4 x \sqrt{-g}\,L(N,K, {\cal S}, {\cal R}, {\cal U};t) 
\nonumber \\
& &+\int d^4 x \sqrt{-g}\,L_m (g_{\mu \nu},\Psi_m)\,,
\label{action}
\ea
where $L_m$ is the matter Lagrangian, and
\ba
L  &=&
A_2(N,t)+A_3(N,t)K \nonumber \\
& &
+A_4(N,t) (K^2-{\cal S}) +B_4(N,t){\cal R} 
\nonumber \\
& &
+A_5(N,t) K_3 
+B_5(N,t) \left( {\cal U}-\frac12 K {\cal R} \right)\,.
\label{LGLPV}
\ea
Here we have defined $K \equiv {K^{\mu}}_{\mu}$, 
${\cal S} \equiv K_{\mu \nu} K^{\mu \nu}$,
${\cal R} \equiv
{{\cal R}^{\mu}}_{\mu}$, 
${\cal U} \equiv {\cal R}_{\mu \nu} K^{\mu \nu}$, 
and $K_3 \equiv K^3-3KK_{\mu \nu}K^{\mu \nu}
+2K_{\mu \nu}K^{\mu \lambda}{K^{\nu}}_{\lambda}$ 
with arbitrary functions $A_{2,3,4,5}$ and $B_{4,5}$ 
that depend on $N$ and $t$ \cite{Gleyzes13,GLPV}.

The coefficients $G_2,G_3,G_4,G_5$ in Horndeski theories are related to
$A_2,A_3,A_4,A_5$ and $B_{4}, B_{5}$, as \cite{Gleyzes13}
\ba
& & A_2=G_2-XF_{3,\phi}\,,\label{A2}\nonumber \\
& & A_3=2(-X)^{3/2}F_{3,X}-2\sqrt{-X}G_{4,\phi}\,,
\label{A3} \nonumber \\
& & A_4=-G_4+2XG_{4,X}+XG_{5,\phi}/2\,,
\label{A4} \nonumber \\
& & B_4=G_4+X(G_{5,\phi}-F_{5,\phi})/2\,,
\label{B4}  \nonumber \\
& & A_5=-(-X)^{3/2}G_{5,X}/3\,, \nonumber \\
& & B_5=-\sqrt{-X}F_{5}\,,
\label{AB}
\ea
where a comma in a lower index represents a partial derivative
with respect to a given scalar quantity, and 
$F_3$ and $F_5$ are auxiliary functions satisfying
$G_3=F_3+2XF_{3,X}$ and $G_{5,X}=F_5/(2X)+F_{5,X}$.
{}From these relations it follows that Horndeski theories 
obey the two conditions
\be
A_4 = 2XB_{4,X}-B_4\,,\qquad
A_5 =-\frac13 XB_{5,X}\,.
\label{ABcon}
\ee

GLPV theories are described by the Lagrangian (\ref{LGLPV}) 
without imposing the two constraints (\ref{ABcon}). 
Even without these restrictions, the linear perturbation equations 
of motion on the flat FLRW background remain of second order
without having an extra propagating scalar degree 
of freedom \cite{GLPV,Hami,Hami2}.

For the study of growth of large-scale structures,
we have taken into account the matter Lagrangian $L_m$ 
in Eq.~(\ref{action}).
We assume that, in the frame described by the metric 
$g_{\mu \nu}$ (dubbed the Jordan frame), there is no explicit 
coupling between the scalar field $\phi$ and matter fields $\Psi_m$.
The matter energy-momentum tensor following from $L_m$ 
is given by 
\be
T^{\mu \nu}=\frac{2}{\sqrt{-g}}
\frac{\delta(\sqrt{-g}\,L_m)}{\delta g_{\mu \nu}}\,.
\label{Tmunu}
\ee
In what follows we describe the matter component as a 
barotropic perfect fluid, whose background energy density and pressure 
are given, respectively, by $\rho$ and $P$.

\section{Linear perturbations}
\label{persec}

The expansion of the action (\ref{action}) up to second order in 
the perturbations on the flat FLRW background gives rise to
the background and linear perturbation equations of motion.
This was already derived in Refs.~\cite{Gleyzes13,Gergely,DKT}, 
so we simply quote the results in the following. 
We consider four scalar metric perturbations $A$, $\psi$, 
$\zeta$, $E$ and tensor perturbations $\gamma_{ij}$ 
described by the line element 
\ba
\hspace{-0.3cm}
ds^2 &=& 
-(1+2A)dt^2+2 \partial_{i} \psi dt dx^i \nonumber \\
\hspace{-0.3cm}
& &+a^2(t) [ (1+2\zeta) \delta_{ij} 
+2\partial_i \partial_j E +\gamma_{ij} ]dx^i dx^j\,,
\label{permet}
\ea
and choose the unitary gauge  
\be
\delta \phi=0\,,\qquad E=0\,,
\ee
where $\delta \phi$ is the perturbation of $\phi$.

The background values (represented by an overbar) 
of the extrinsic and intrinsic curvatures are given, respectively, by 
$\bar{K}_{\mu \nu}=H\bar{h}_{\mu \nu}$ and
$\bar{{\cal R}}_{\mu \nu}=0$, where $H=\dot{a}/a$ is 
the Hubble parameter. Then, it follows that 
$\bar{K}=3H$,  $\bar{\cal S}=3H^{2}$, and 
$\bar{{\cal R}}=\bar{\cal U}=0$.
We consider the perturbations of these geometric scalars, e.g., 
$\delta N=N-1$, $\delta K=K-3H$.
The perturbations of the matter energy-momentum tensors 
are denoted as 
\be
\delta T^0_0=-\delta \rho\,,\qquad
\delta T^0_i=\partial_i \delta q\,,\qquad
\delta T^{i}_{j}=\delta P \delta^{i}_{j}\,,
\ee
where Latin indices correspond to components in a 
three-dimensional space-adapted basis.

Expanding the action (\ref{action}) up to first order 
in scalar perturbations and varying the first-order action 
with respect to $\delta N$ and $\delta a$, 
we obtain the background equations
\ba
& &
\bar{L}+L_{,N}-3H {\cal F}=\rho\,,
\label{back1}\\
& &
\bar{L}-\dot{\cal F}-3H{\cal F}=-P\,,
\label{back2}
\ea
respectively, 
where ${\cal F} \equiv L_{,K}+2HL_{,\cal S}$. 
The matter component obeys the continuity equation 
\be
\dot{\rho}+3H (\rho+P)=0\,.
\ee

The second-order action for tensor perturbations 
derived from Eq.~(\ref{action}) reads 
\be
S_2^{(h)}=\int d^4 x\,a^3 q_{\rm t} \delta^{ik} \delta^{jl}
\left( \dot{\gamma}_{ij} \dot{\gamma}_{kl}
-\frac{c_{\rm t}^2}{a^2}\partial \gamma_{ij} 
\partial \gamma_{kl} \right)\,, 
\label{L2ten}
\ee
where
\be
q_{\rm t} \equiv \frac{L_{,\cal S}}{4}\,,\qquad
c_{\rm t}^2 \equiv \frac{{\cal E}}{L_{,\cal S}}\,,
\label{qtctdef}
\ee
and
\be
{\cal E} \equiv L_{,{\cal R}}+\frac12 \dot{L}_{,{\cal U}}
+\frac32 H L_{,{\cal U}}\,.
\ee
We require the two conditions 
\ba
& & q_{\rm t}>0\,,
\label{tenghost} \\
& & c_{\rm t}^2>0\,,
\label{tensta}
\ea
to avoid the tensor ghost and small-scale 
Laplacian instabilities, respectively.

In the presence of matter, the second-order action $S_2^{(s)}$ 
for scalar perturbations in GLPV theories was given 
in Ref.~\cite{Gergely,Hami2,DKT}. 
GLPV theories satisfy conditions for the absence of spatial 
derivatives higher than second order \cite{GLPV,Kasecosmo}.
Varying $S_2^{(s)}$ with respect to $\delta N$, $\partial^2 \psi$, 
and $\zeta$, it follows that 
\ba
& &
\left( 2L_{,N}+L_{,NN}-6H{\cal W}+12L_{,{\cal S}}H^2 \right) \delta N 
\nonumber \\
& &+\left( 3\dot{\zeta}-\frac{\partial^2 \psi}{a^2} \right){\cal W}
-4({\cal D}+{\cal E})\frac{\partial^2 \zeta}{a^2}=\delta \rho\,,
\label{pereq1} \\
& &
{\cal W} \delta N-4L_{,{\cal S}} \dot{\zeta}=-\delta q\,,
\label{pereq2} \\
& &
\frac{1}{a^3} \frac{d}{dt} \left( a^3 {\cal Y} \right)
+4({\cal D}+{\cal E}) \frac{\partial^2 \delta N}{a^2}
+\frac{4{\cal E}}{a^2} \partial^2 \zeta \nonumber \\
& &-3(\rho+P) \delta N
=3\delta P\,,
\label{pereq3} 
\ea
respectively, where 
\ba
{\cal W} &\equiv& L_{,KN}+2HL_{,{\cal S}N}+4L_{,{\cal S}}H\,,\\
{\cal D} &\equiv& L_{,N{\cal R}}-\frac12 \dot{L}_{,{\cal U}}
+HL_{,N{\cal U}}\,,\\
{\cal Y} &\equiv& 4L_{,\cal S} \frac{\partial^2 \psi}{a^2}
-3\delta q\,.
\label{calY}
\ea
The continuity equations ${\delta T^{\mu}}_{0;\mu}=0$ and 
${\delta T^{\mu}}_{i;\mu}=0$ lead, respectively, to 
\ba
& &
\dot{\delta \rho}+3H \left( \delta \rho+\delta P \right)
\nonumber \\
& &
=-(\rho+P) \left( 3\dot{\zeta}-\frac{\partial^2 \psi}{a^2} \right)
-\frac{\partial^2 \delta q}{a^2}\,,
\label{mper1} \\
& &
\dot{\delta q}+3H \delta q=-(\rho+P)\delta N-\delta P\,.
\label{mper2}
\ea
Substituting Eq.~(\ref{calY}) into Eq.~(\ref{pereq3}) and 
employing Eq.~(\ref{mper2}), we obtain
\be
( \dot{L}_{,\cal S}+HL_{,\cal S} )\psi
+L_{,\cal S} \dot{\psi}+( {\cal D}+{\cal E} ) \delta N
+{\cal E} \zeta=0\,.
\label{pereq4}
\ee

On using Eqs.~(\ref{pereq1}) and (\ref{pereq2}), the second-order 
scalar action $S_2^{(s)}$ can be expressed in terms of $\zeta$, 
the matter perturbations, and their derivatives \cite{Gergely}. 
Provided that the matter component does not correspond to 
a ghost mode, the scalar ghost is absent under 
the condition \cite{Gleyzes13,Gergely,Kasecosmo}
\be
q_{\rm s}=\frac{8q_{\rm t} (16q_{\rm t}w_s+3{\cal W}^2)}
{{\cal W}^2}>0\,,
\label{scaghost}
\ee
where $w_s \equiv 2L_{,N}+L_{,NN}-6H{\cal W}+12H^2L_{,\cal S}$. 
Under the condition (\ref{tenghost}), Eq.~(\ref{scaghost}) 
translates to $16q_{\rm t}w_s+3{\cal W}^2>0$.

In GLPV theories the scalar propagation speed squared $c_{\rm s}^2$
is affected by the presence of 
matter \cite{Gergely,GLPV,DKT}.
For nonrelativistic matter characterized 
by $P=0$ and $\delta P=0$, the value of $c_{\rm s}^2$ in the 
small-scale limit reads
\be
c_{\rm s}^2=\frac{2}{q_{\rm s}} \left[ \dot{\cal M}+H{\cal M}
-{\cal E}-\frac{4L_{,\cal S}^2\,\rho}{{\cal W}^2} 
\left( 1+2\alpha_{\rm H} \right) \right]\,,
\label{cs2}
\ee
where 
\ba
{\cal M} &\equiv& \frac{4L_{,\cal S} ({\cal D}+{\cal E})}
{{\cal W}}\,,\\
\alpha_{\rm H} &\equiv& 
\frac{{\cal D}+{\cal E}}{L_{,\cal S}}-1
=c_{\rm t}^2-1+\frac{{\cal D}}{L_{,\cal S}}\,.
\ea
The parameter $\alpha_{\rm H}$ characterizes the 
deviation from Horndeski theories \cite{Hami2}. 
To avoid the small-scale instability of scalar perturbations,  
we require that
\be
c_{\rm s}^2>0\,.
\label{scasta}
\ee
The four conditions (\ref{tenghost}), (\ref{tensta}), (\ref{scaghost}), 
and (\ref{scasta}) need to be satisfied for theoretical consistency.

We define the gauge-invariant gravitational potentials \cite{Bardeen}
\be
\Psi \equiv \delta N+\dot{\psi}\,,\qquad
\Phi \equiv \zeta+H \psi\,,
\label{Psidef}
\ee
and the gravitational slip parameter
\be
\eta \equiv -\frac{\Phi}{\Psi}\,.
\label{etadef}
\ee
Then, Eq.~(\ref{pereq4}) can be written as
\be
\Psi+\Phi=\frac{\dot{q}_{\rm t}}{Hq_{\rm t}} \left( \zeta-\Phi \right)
-\left( c_{\rm t}^2 -1 \right)\zeta-\alpha_{\rm H} \delta N\,.
\ee
This shows that the deviation of $\eta$ from 1 is induced by the variation 
of $q_{\rm t}$, the deviation of $c_{\rm t}^2$ from 1, 
and the deviation parameter $\alpha_{\rm H}$ from Horndeski theories.

We also introduce the effective gravitational potential 
\be
\Phi_{\Sigma} \equiv \frac12 (\Psi-\Phi)\,,
\label{PhiWL}
\ee
which is associated with the deviation of light rays in weak lensing 
and CMB observations \cite{lensing}.

\section{Subhorizon perturbations}
\label{subsec}

To study the growth of structures during the matter-dominated epoch, 
we shall take into account nonrelativistic matter satisfying 
$P=0$ and $\delta P=0$ 
for the Lagrangian $L_m$. 
We also define the gauge-invariant matter perturbation 
\be
\delta_m \equiv \delta -3H v\,,
\ee
where $\delta \equiv \delta \rho/\rho$ and $v \equiv \delta q/\rho$.
Taking the time derivative of Eq.~(\ref{mper1}) in Fourier space 
and using Eq.~(\ref{mper2}), we obtain 
\be
\ddot{\delta}_m+2H \dot{\delta}_m+\frac{k^2}{a^2} \Psi
=-3\ddot{B}-6H\dot{B}\,,
\label{delmeq}
\ee
where $k$ is a comoving wave number 
and $B \equiv \zeta+Hv$. 
The gravitational potential $\Psi$ works as a source term 
for the growth of matter perturbations.

To estimate the evolution of $\Psi$, we consider the perturbations 
deep inside the Hubble radius, i.e., $k/a \gg H$.
In Fourier space we employ the subhorizon approximation 
under which the dominant contributions to the perturbation 
equations are the terms involving $\delta \rho$ and the terms 
multiplied by $k^2/a^2$ \cite{subap}. 
In GR, the accuracy of this approximation was 
numerically confirmed for the subhorizon 
perturbations \cite{Caldwell}.

By employing the subhorizon approximation in modified gravity theories, 
we consider theories in which the deviation from GR 
is not so significant in a way that the terms other than those 
containing $k^2/a^2$ and $\delta \rho$ 
are still subdominant to the perturbation 
equations, as in the case of GR. 
For example, the orders of the terms 
$L_{,N}\delta N$ and $L_{,{\cal S}}H^2\delta N$
in Eq.~(\ref{pereq1}) are regarded to be at most of 
the orders of $M_{\rm pl}^2 H^2 \delta N$. 
Under the subhorizon approximation, Eq.~(\ref{pereq1}) reads
\be
\left( \frac{k}{aH} \right)^2 \left[ \alpha_{\W} \chi
+4(1+\alpha_{\rm H}) \zeta \right] \simeq 
6\Omega_m \delta\,,
\label{sub1}
\ee
where 
\be
\alpha_{\W} \equiv \frac{{\cal W}}{HL_{,\cal S}}\,,
\qquad 
\chi \equiv H \psi\,,\qquad
\Omega_m \equiv \frac{\rho}{6H^2 L_{,{\cal S}}}\,.
\label{Omem}
\ee
The perturbation $\chi$ is related to the gravitational potential 
$\Phi$, as $\Phi=\zeta+\chi$.
The definition of $\Omega_m$ comes from the fact that 
the Friedmann equation (\ref{back1}) can be written as 
\ba
6H^2L_{,{\cal S}}
&=& \rho-A_2-6H^3 A_5-A_{2,N}-3HA_{3,N}
\nonumber \\
& &+6H^2 L_{,{\cal S}N}+12H^3 A_{5,N}\,,
\label{Fri}
\ea
where we used the relation $A_4=-L_{,{\cal S}}-3HA_5$. 
Alternatively, one may introduce the General-Relativistic form 
of the matter density parameter $\tilde{\Omega}_m=\rho/(3H^2M_{\rm pl}^2)$ 
by expressing Eq.~(\ref{Fri}) in the form $3H^2M_{\rm pl}^2=\rho+\cdots$, 
where $M_{\rm pl}$ is the reduced Planck mass.
In this case, the rhs of Eq.~(\ref{sub1}) is replaced by 
$3M_{\rm pl}^2 \tilde{\Omega}_m \delta/L_{,{\cal S}}$.

Expressing Eq.~(\ref{pereq4}) in terms of $\Psi$, $\zeta$, $\chi$, and 
$\dot{\chi}$, it follows that 
\be
(1+\alpha_{\rm H}) \Psi+c_{\rm t}^2 \zeta
+(1+\epsilon_{\S}-\alpha_{\rm H} \epsilon_{H})\chi
=\alpha_{\rm H} \frac{\dot{\chi}}{H}\,,
\label{sub2}
\ee
where
\be
\epsilon_{\S} \equiv \frac{\dot{L}_{,\cal S}}{HL_{,\cal S}}\,,\qquad
\epsilon_{H} \equiv -\frac{\dot{H}}{H^2}\,.
\ee

Under the subhorizon approximation
the continuity equation (\ref{mper1}) reads
\be
\dot{\delta \rho}+3H \delta \rho \simeq 
\frac{k^2}{a^2} \left( -\rho \psi+4L_{,\cal S} \dot{\zeta}
-{\cal W} \delta N \right)\,,
\label{submiddle}
\ee
where we employed Eq.~(\ref{pereq2}) and ignored the term $3\dot{\zeta}$ 
relative to $(k^2/a^2)\psi$.
Taking the time derivative of Eq.~(\ref{sub1}), we 
can eliminate the terms $\dot{\delta \rho}$ and $\delta \rho$ 
on the lhs of Eq.~(\ref{submiddle}). 
This process leads to 
\ba
& &
\alpha_{\W}\Psi+\left( \alpha_{\W}+\epsilon_{\W} \alpha_{\W}
+6\Omega_m \right)\chi \nonumber \\
& &
+4 \left[ (1+\alpha_{\rm H})(1+\epsilon_{\S})+\epsilon_{\alpha_{\rm H}}
 \right]\zeta \simeq -4\alpha_{\rm H} \frac{\dot{\zeta}}{H}\,,
\label{sub3}
\ea
where 
\be
\epsilon_{\W} \equiv \frac{\dot{\cal W}}{H{\cal W}}\,,\qquad
\epsilon_{\alpha_{\rm H}} \equiv 
\frac{\dot{\alpha}_{\rm H}}{H}\,.
\ee

In the unitary gauge ($\delta \phi=0$),  the mass $M$ of 
the scalar degree of freedom does not explicitly appear in the 
perturbation equations of motion.
The above subhorizon approximation is valid for $M$ 
smaller than $c_{\rm s}k/a$ \cite{DKT11,Amendola,Bellini}.
In the regime $M \gg c_{\rm s}k/a$, the scalar field is nearly 
frozen to recover the General-Relativistic behavior, 
so that $G_{\rm eff}$ is very close to $G$. 
In fact, this happens for viable $f(R)$ dark 
energy models in the early stage of the 
matter era \cite{frper}.

Since $\alpha_{\rm H}=0$ in Horndeski theories, the terms 
on the rhs of Eqs.~(\ref{sub2}) and (\ref{sub3}) vanish.
In this case, we can express $\chi$ and $\zeta$ as a function of $\Psi$ from 
Eqs.~(\ref{sub2}) and (\ref{sub3}) and then derive $\Psi$
in terms of $\delta$ by using Eq.~(\ref{sub1}).

In GLPV theories we need to deal with Eqs.~(\ref{sub2}) and (\ref{sub3})
as the differential equations involving $\dot{\chi}$ and $\dot{\zeta}$.
We define the quantities
\be
\epsilon_{\zeta} \equiv \frac{\dot{\zeta}}{H\zeta}\,,\qquad
\epsilon_{\chi} \equiv \frac{\dot{\chi}}{H\chi}\,.
\label{epdef}
\ee
If $\epsilon_{\zeta}$ and $\epsilon_{\chi}$
are smaller than the order of 1, the terms on the rhs of 
Eqs.~(\ref{sub2}) and (\ref{sub3}) can be regarded as 
the corrections to those on the lhs.
Let us consider this situation and express 
$\dot{\zeta}$ and $\dot{\chi}$ in terms of $\epsilon_{\zeta}$ 
and $\epsilon_{\chi}$.
Of course the evolution of $\epsilon_{\zeta}$ 
and $\epsilon_{\chi}$ is known only by solving the full perturbation equations 
for a given theory, so Eqs.~(\ref{sub1}), (\ref{sub2}), 
and (\ref{sub3}) are not closed for $\alpha_{\rm H} \neq 0$. 

We define the effective gravitational coupling 
$G_{\rm eff}$, as
\be
\frac{k^2}{a^2}\Psi=-4\pi G_{\rm eff} \rho\,\delta_m\,,
\label{Geffdef}
\ee
which works as a source term for the growth 
of $\delta_m$ on the lhs of Eq.~(\ref{delmeq}).
The potential (\ref{PhiWL}) obeys
\be
\frac{k^2}{a^2}\Phi_{\Sigma}=-4\pi G \Sigma \rho\,\delta_m\,,
\label{PhiSigma}
\ee
where 
\be
\Sigma=\frac{1+\eta}{2} \frac{G_{\rm eff}}{G}\,.
\label{Sigma}
\ee
Recall that the gravitational slip parameter is given by 
$\eta=-(\zeta+\chi)/\Psi$.

It is convenient to express $G_{\rm eff}$, $\eta$, and $\Sigma$ 
in terms of the quantities $q_{\rm t}$, $c_{\rm t}^2$, 
$q_{\rm s}$, and $c_{\rm s}^2$ because the signs of them 
need to be positive.
Substituting the relations ${\cal E}=4q_{\rm t}c_{\rm t}^2$, 
${\cal D}+{\cal E}=4q_{\rm t}(1+\alpha_{\rm H})$,  
${\cal W}=4Hq_{\rm t}\alpha_{\W}$, and 
${\cal M}=16q_{\rm t}(1+\alpha_{\rm H})/(H\alpha_{\W})$ 
into Eq.~(\ref{cs2}), it follows that 
\ba
\hspace{-0.7cm}
c_{\rm s}^2&=&\frac{32q_{\rm t}(1+\alpha_{\rm H})}
{q_{\rm s} \alpha_{\W}} \biggl[ 1+2\epsilon_{\S} 
-\epsilon_{\W}+\frac{\epsilon_{\alpha_{\rm H}}}{1+\alpha_{\rm H}} 
\nonumber \\
\hspace{-0.7cm}
& &~~~~~~~~~~~~~~~~~~
-\frac{c_{\rm t}^2 \alpha_{\W}}{4(1+\alpha_{\rm H})} 
-\frac{6\Omega_m}{\alpha_{\W}} 
\frac{1+2\alpha_{\rm H}}{1+\alpha_{\rm H}} \biggr].
\label{cs}
\ea
This relation can be employed to express $\epsilon_{\W}$
in terms of $c_{\rm s}^2$. 
On using Eqs.~(\ref{sub1}), (\ref{sub2}), and (\ref{sub3}) 
with Eq.~(\ref{epdef}) and the relation $\delta_m \simeq \delta$ 
under the subhorizon approximation, we obtain
\ba
\frac{G_{\rm eff}}{G} 
&=& \frac{M_{\rm pl}^2}{8q_{\rm t}} 
\frac{f_1}{f_2}\,, \label{quasi1} \\
\eta 
&=& \frac{f_3}{f_1}\,,\label{quasi2} \\
\Sigma 
&=& \frac{f_1+f_3}{2f_2} \frac{M_{\rm pl}^2}
{8q_{\rm t}}\,,
\label{quasi3} 
\ea
where $G=(8\pi M_{\rm pl}^2)^{-1}$ is the 
gravitational constant, and 
\begin{widetext}
\ba
f_1 &=&
64 [2 (1+\alpha_{\rm H})(1-\alpha_{\rm H} \epsilon_{H}
-\alpha_{\rm H} \epsilon_{\chi}
+\epsilon_{\S})(1+\epsilon_{\S}+\alpha_{\rm H}+
\alpha_{\rm H}\epsilon_{\S}+\alpha_{\rm H}\epsilon_{\zeta}
+\epsilon_{\alpha_{\rm H}})
+3\alpha_{\rm H} c_{\rm t}^2 \Omega_m ] q_{\rm t} \nonumber \\
&&
-32c_{\rm t}^2  \alpha_{\W} \left[ 2+\epsilon_{\alpha_{\rm H}}
+2\epsilon_{\S}+2\alpha_{\rm H} (1+\epsilon_{\S}) \right] q_{\rm t} 
+\alpha_{\W}^2 c_{\rm t}^2 (q_{\rm s}c_{\rm s}^2
+8q_{\rm t}c_{\rm t}^2)\,,
\label{f1} \\
f_2 &=& 
(1+\alpha_{\rm H})\{(1+\alpha_{\rm H})q_{\rm s}c_{\rm s}^2 
\alpha_{\W}^2+8\alpha_{\rm H} 
[ 24(1+\alpha_{\rm H}) \Omega_m+c_{\rm t}^2 \alpha_{\W}^2
-4(1+\alpha_{\rm H}) \alpha_{\W} (1+\epsilon_{H}+\epsilon_{\S}
+\epsilon_{\chi}-\epsilon_{\zeta})]q_{\rm t}\}\,,\label{f2} \\
f_3 &=&
(1+\alpha_{\rm H})\{ 
64[ 2(1+\alpha_{\rm H})(1+\epsilon_{\S}+\alpha_{\rm H}+
\alpha_{\rm H}\epsilon_{\S}+\alpha_{\rm H}\epsilon_{\zeta}
+\epsilon_{\alpha_{\rm H}})
+3\alpha_{\rm H} \Omega_m]q_{\rm t} \nonumber \\
& &
-32\alpha_{\W} [1+c_{\rm t}^2+\epsilon_{\alpha_{\rm H}}
+\epsilon_{\S}+\alpha_{\rm H} (2+\epsilon_{H}+2\epsilon_{\S}+\epsilon_{\chi})]q_{\rm t}
+\alpha_{\W}^2 (q_{\rm s}c_{\rm s}^2
+8q_{\rm t}c_{\rm t}^2)\}\,.
\label{f3}
\ea
\end{widetext}

Equation (\ref{quasi1}) shows that $G_{\rm eff}$ is scale independent. 
This comes from the fact that we have ignored the field 
mass $M$ relative to $c_{\rm s}k/a$ for its derivation.
In other words, the results (\ref{quasi1})-(\ref{quasi3}) are 
valid in the regime $M \ll  c_{\rm s}k/a$. 
For the scalar degree of freedom associated 
with dark energy, the field mass is usually smaller than $c_{\rm s}k/a$ 
at the late cosmological epoch.
For some dark energy models in which the chameleon 
mechanism \cite{chame} is at work in the region of high density, 
there is a transition from the General-Relativistic regime 
($M \gg c_{\rm s}k/a$) with $G_{\rm eff} \simeq G$ to the 
scalar-tensor regime ($M \ll c_{\rm s}k/a$) with $G_{\rm eff}$ 
given by Eq.~(\ref{quasi1}) \cite{frper,tran}.
For the model discussed later in Sec.~\ref{obsersec}, 
the condition $M \ll  c_{\rm s}k/a$
is satisfied for subhorizon perturbations.

In Horndeski theories ($\alpha_{\rm H}=0$, $\epsilon_{\alpha_{\rm H}}=0$), 
the above analytic solutions are closed.
In GLPV theories the terms $\alpha_{\rm H}\epsilon_{\zeta}$ 
and $\alpha_{\rm H}\epsilon_{\chi}$ do not vanish 
in Eqs.~(\ref{f1})-(\ref{f3}). 
In this case, we can check the validity of the subhorizon approximation 
by solving the full perturbation equations of motion numerically 
for a given theory and by comparing the full results with the 
estimations (\ref{quasi1})-(\ref{quasi3}) derived after the substitution of numerical 
values of $\epsilon_{\zeta}$ and $\epsilon_{\chi}$ into 
Eqs.~(\ref{f1})-(\ref{f3}). 
In Sec.~\ref{obsersec} we shall do so for a concrete theory 
in the framework of GLPV theories.

\section{Possibility of realizing weak gravity 
in the cosmic growth history}
\label{weaksec}

In this section we discuss the possibility of realizing a gravitational interaction 
weaker than that in GR ($G_{\rm eff}<G$) on scales 
relevant to large-scale structures. 
We shall focus on GLPV theories described by the Lagrangian (\ref{LGLPV}).
Then, the tensor propagation speed squared is given by 
\be
c_{\rm t}^2=\frac{B_4+\dot{B}_5/2}{-A_4-3HA_5}\,.
\label{ct2}
\ee
For the evaluation of the scalar propagation speed squared (\ref{cs}), 
it is convenient to express $\Omega_m$ by using the background 
equations of motion (\ref{back1}) and (\ref{back2}), i.e., 
\be
\frac{6\Omega_m}{\alpha_{\W}}
=\frac{L_{,N}+\dot{\cal F}}{H{\cal W}}\,.
\label{Omere}
\ee

In what follows we shall discuss the cases of Horndeski 
and GLPV theories separately. 

\subsection{Horndeski theories}

Substituting $\alpha_{\rm H}=0$ and $\epsilon_{\alpha_{\rm H}}=0$ 
into Eqs.~(\ref{f1})-(\ref{f3}), we obtain 
$f_1=\alpha_{\W}^2 q_{\rm s} c_{\rm s}^2 c_{\rm t}^2
+8(\alpha_{\W}c_{\rm t}^2-4-4\epsilon_{\S})^2q_{\rm t}$, 
$f_2=\alpha_{\W}^2q_{\rm s}c_{\rm s}^2$, and 
$f_3=\alpha_{\W}^2q_{\rm s}c_{\rm s}^2+
8(\alpha_{\W}-4)(\alpha_{\W}c_{\rm t}^2-4-4\epsilon_{\S})q_{\rm t}$. 
Then, Eqs.~(\ref{quasi1})-(\ref{quasi3}) reduce to 
\ba
\hspace{-0.2cm}
\frac{G_{\rm eff}}{G}
&=& \frac{M_{\rm pl}^2c_{\rm t}^2}{8q_{\rm t}} 
\left( 1+\frac{8Q^2}
{\alpha_{\W}^2 c_{\rm s}^2 c_{\rm t}^2} \frac{q_{\rm t}}{q_{\rm s}}
\right)\,,\label{GeffHo} \\
\hspace{-0.2cm}
\eta 
&=& \frac{\alpha_{\W}^2q_{\rm s}c_{\rm s}^2+
8(\alpha_{\W}-4)Q q_{\rm t}}
{\alpha_{\W}^2 q_{\rm s} c_{\rm s}^2 c_{\rm t}^2
+8Q^2q_{\rm t}}\,,\label{etaHo} \\
\hspace{-0.2cm}
\Sigma 
&=& \frac{M_{\rm pl}^2(1+c_{\rm t}^2)}{16q_{\rm t}} 
\left[ 1+\frac{8Q (Q+\alpha_{\W}-4)}
{\alpha_{\W}^2c_{\rm s}^2(1+c_{\rm t}^2)} 
\frac{q_{\rm t}}{q_{\rm s}} \right]\,,
\label{SigmaHo}
\ea
where 
\be
Q \equiv \alpha_{\W}c_{\rm t}^2-4-4\epsilon_{\S}\,.
\label{beta}
\ee

Provided the rhs of Eq.~(\ref{GeffHo}) is smaller than 1, 
it follows that $G_{\rm eff}<G$. The contribution 
$M_{\rm pl}^2c_{\rm t}^2/(8q_{\rm t})$ in $G_{\rm eff}/G$
originates from the tensor part, whereas the second term in the bracket 
of Eq.~(\ref{GeffHo}) comes from the interaction between the scalar field 
and matter. Under the no-ghost and stability requirements (\ref{tenghost}), 
(\ref{tensta}), (\ref{scaghost}), and (\ref{scasta}), the latter contribution 
is always positive.
Hence the {\it necessary} condition for realizing a gravitational 
interaction weaker than that in GR reads
\be
\frac{M_{\rm pl}^2c_{\rm t}^2}{8q_{\rm t}} <1\,.
\label{ctcon}
\ee
Due to the presence of the scalar-matter interaction, the condition 
(\ref{ctcon}) is not sufficient for realizing $G_{\rm eff}<G$.

The quantities $q_{\rm t}$ and $c_{\rm t}^2$ 
are given, respectively, by 
\ba
q_{\rm t} &=& -\frac{1}{4} \left( A_4+3HA_5 \right)\,,\\
c_{\rm t}^2 &=& 
\frac{B_4+\dot{B}_5/2}{B_4-2XB_{4,X}+HXB_{5,X}}\,,
\label{ctHo}
\ea
where we used the two relations (\ref{ABcon}).  
The explicit form of the quantity $Q$ in Eq.~(\ref{beta}) 
is given by 
\begin{widetext}
\be
Q=\frac{A_{3,N}c_{\rm t}^2
+4H A_{4,N}c_{\rm t}^2+4\dot{A}_4
+4H A_4 (1-c_{\rm t}^2)+6H^2 A_{5,N}c_{\rm t}^2
+12H \dot{A}_5+12[H^2(1-c_{\rm t}^2)+\dot{H}]A_5}
{H(-A_4-3HA_5)}\,. 
\label{betaHo}
\ee
\end{widetext}
In GR we have $-A_4=B_4=M_{\rm pl}^2/2$,
$A_3=0$, and $A_5=B_5=0$, in which case 
$q_{\rm t}=M_{\rm pl}^2/8$, $c_{\rm t}^2=1$ and $Q=0$.
In modified gravitational theories, we generally have 
$M_{\rm pl}^2c_{\rm t}^2/(8q_{\rm t}) \neq 1$ and $Q \neq 0$.

As an example, let us consider theories
described by the Lagrangian 
\be
L=A_2(N,t)+A_3(N,t)K+A_4(t) \left( K^2-{\cal S}
-{\cal R} \right)\,,
\label{L1}
\ee
in which case $q_{\rm t}=-A_4/4$, $c_{\rm t}^2=1$, 
and $Q=-(A_{3,N}+4\dot{A}_4)/(HA_4)$. 
The theory given by $L=-\epsilon(\phi)X/2-V(\phi)+M_{\rm pl}^2F(\phi)R/2$, 
where $\epsilon$, $V$, and $F$ are arbitrary functions of $\phi$, 
belongs to a subclass of Eq.~(\ref{L1}). 
For the choices $\epsilon(\phi)=\omega_{\rm BD}M_{\rm pl}/\phi$ 
and $F(\phi)=\phi/M_{\rm pl}$, this Lagrangian
encompasses Brans-Dicke theory \cite{Brans} as a special case.
On using the correspondence (\ref{B4}), the Lagrangian of 
this theory is equivalent to Eq.~(\ref{L1}) with the functions 
\ba
A_2 &=&-\frac12 \epsilon(\phi)X-V(\phi)\,,\nonumber \\
A_3 &=&  -\frac{1}{N}M_{\rm pl}^2 
\dot{\phi}F_{,\phi}(\phi)\,,\nonumber \\
A_4 &=& -B_4=-\frac12 M_{\rm pl}^2F(\phi)\,.
\label{L2}
\ea

In this case the relation 
${\cal W}=-{\cal F}=M_{\rm pl}^2 (2HF+\dot{\phi}F_{,\phi})$ 
holds, so Eq.~(\ref{cs}) reduces to 
$c_{\rm s}^2=1$ by using Eq.~(\ref{Omere}).
{}From Eqs.~(\ref{GeffHo})-(\ref{SigmaHo}) it follows that 
\ba
G_{\rm eff}
&=& \frac{G}{F} \left( 1+ 
\frac{M_{\rm pl}^2 F_{,\phi}^2}
{2F\epsilon+3M_{\rm pl}^2 F_{,\phi}^2} \right)\,,
\label{Geffex1}\\
\eta
&=&
\frac{F\epsilon+M_{\rm pl}^2 F_{,\phi}^2}
{F\epsilon+2M_{\rm pl}^2 F_{,\phi}^2}\,,\label{etaex1}\\
\Sigma &=& 
\frac{1}{F}\,.
\label{Sigex1}
\ea
The conditions $q_{\rm t}>0$ and $q_{\rm s}>0$ 
translate to $F>0$ and $2F\epsilon+3M_{\rm pl}^2 F_{,\phi}^2>0$, 
respectively, so $G_{\rm eff}$ is larger than $G/F$ due to the presence of
the second term in the bracket of Eq.~(\ref{Geffex1}).
{}From Eq.~(\ref{etaex1}) the parameter $\eta$ 
is smaller than $1$.
The enhancement of $G_{\rm eff}$
is compensated by the smallness of $\eta$, so that 
we obtain the value (\ref{Sigex1}). 
In other words, we have 
$Q+\alpha_{\W}-4=0$ in Eq.~(\ref{SigmaHo})
for the model discussed above.

Let us consider the full Lagrangian (\ref{LGLPV}) with 
the Horndeski relations (\ref{ABcon}). The behavior of the quantities 
$q_{\rm t}$ and $c_{\rm t}^2$ is crucial for the realization 
of the condition (\ref{ctcon}). 
The quantity $q_{\rm t}=L_{,\cal S}/4$ is associated with the 
matter density parameter $\Omega_m$ defined in Eq.~(\ref{Omem}), 
as $\Omega_m=\rho/(24H^2 q_{\rm t})$.
The matter perturbation equation (\ref{delmeq}) can be expressed 
by using $\Omega_m$, as
\be
\delta_m''+ \left( 2+\frac{H'}{H} \right) \delta_m'
-\frac32 c_{\rm t}^2 \left( 1+\frac{8Q^2}
{\alpha_{\W}^2 c_{\rm s}^2 c_{\rm t}^2} 
\frac{q_{\rm t}}{q_{\rm s}}
\right)\Omega_m \delta_m \simeq 0\,,
\label{delmHo}
\ee
where a prime represents a derivative with respect to 
${\cal N}=\ln a$. Note that we neglected the contribution 
on the rhs of Eq.~(\ref{delmeq}) under the 
subhorizon approximation.

In Eq.~(\ref{delmHo}) the quantity $q_{\rm t}$ appearing in 
the denominator of Eq.~(\ref{GeffHo}) 
has been absorbed into the definition of $\Omega_m$.
If $\Omega_m$ is smaller than that in GR due to 
large values of $q_{\rm t}$, it is possible to realize
a cosmic growth rate smaller than that in GR. 
This is one possibility for the realization of weak gravity 
recently studied in Ref.~\cite{Piaweak}. For this purpose 
we require that the scalar-matter coupling $Q$ is suppressed 
to satisfy the condition 
$c_{\rm t}^2 [1+8Q^2q_{\rm t}/(\alpha_{\W}^2c_{\rm s}^2
c_{\rm t}^2q_{\rm s})]\Omega_m<1$.
In dark energy models based on $f(R)$ theories, for example, 
we have that $c_{\rm t}^2=1$, 
$8Q^2q_{\rm t}/(\alpha_{\W}^2c_{\rm s}^2
c_{\rm t}^2q_{\rm s})=1/3$, and that the deviation of
$\Omega_m$ from the value of GR is not so
significant, in which case the growth 
rate of $\delta_m$ is larger than that in GR \cite{frper,tran}.

Let us proceed to the discussion of $c_{\rm t}^2$ appearing 
in Eq.~(\ref{delmHo}).
Due to the relations (\ref{ABcon}), the $X$ dependence in 
$B_4$ and $B_5$ gives rise to the functions $A_4$ and 
$A_5$ depending on $X$ as well as $\phi$.
These terms contribute to the background equations 
of motion as the field energy density and pressure. 
To realize a viable cosmology, 
the derivative terms $2XB_{4,X}$, $HXB_{5,X}$, and 
$\dot{B}_5/2$ in Eq.~(\ref{ctHo}) should be suppressed 
relative to $B_4$ in the deep matter era, 
in which case $c_{\tm}^2$ is close to 1.

If the field $\phi$ is responsible for dark energy, 
the derivative terms mentioned above can be comparable 
to $B_4$ after the onset of cosmic acceleration.
This can lead to a deviation of $c_{\rm t}^2$ from 1. 
Whether $c_{\rm t}^2$ decreases or not depends 
on the models and initial conditions. 
In covariant Galileons \cite{Galileons}, for example, 
the entry into the region $c_{\rm t}^2<1$ can occur 
for late-time tracking solutions \cite{DTPRL}.

{}From Eq.~(\ref{betaHo}) the $t$ and $N$ dependence in
$A_4$ and $A_5$ as well as the $N$ dependence in $A_3$ 
lead to nonzero values of $Q$, which enhances 
the rhs of Eq.~(\ref{GeffHo}). 
Even if $c_{\rm t}^2$ starts to decrease after 
the end of the matter era, the scalar-matter interaction induced 
by the term $Q$ should not be large enough to violate 
the condition $G_{\rm eff}<G$ for the realization of weak gravity.
In covariant Galileons, for example, this scalar-matter interaction 
usually gives rise to a $G_{\rm eff}$ larger than $G$ even 
for $c_{\tm}^2<1$ at the level of linear perturbations \cite{Galimatter}.

To summarize, the modification to $G_{\rm eff}$ coming from 
tensor perturbations is the first crucial factor for realizing 
weak gravity in Horndeski theories, but the condition (\ref{ctcon}) is not sufficient due to the presence of 
the scalar-matter interaction. 
In Sec.~\ref{GLPVsec} we shall study what kind 
of difference arises in GLPV theories.

\subsection{Theories beyond Horndeski}
\label{GLPVsec}

Since GLPV theories are not subject to the 
constraints (\ref{ABcon}), the values of $c_{\tm}^2$ 
are not restricted to be close to 1 even in the early matter era.
In GLPV theories, the general expressions of 
$G_{\rm eff}$, $\eta$, and $\Sigma$ are not as simple as
those in Horndeski theories due to the extra terms  
$\alpha_{\rm H}$ and $\epsilon_{\alpha_{\rm H}}$.
To simplify the analysis, we focus on the theories 
described by the Lagrangian
\be
L=-\frac12 X-V(\phi)-\frac{M_{\rm pl}^2}{2}
(K^2-{\cal S})+\frac{M_{\rm pl}^2}{2}F(\phi){\cal R}\,,
\label{Lagge}
\ee
where $F(\phi)$ is a function of $\phi$ different from 1. 
Unlike Ref.~\cite{DKT}, we do not restrict the situation to the 
case in which $F(\phi)$ is constant. 
In fact, we will show that the theories with a time-varying 
$F(\phi)$ allow for the realization of weak gravity 
in the regime $0<c_{\rm t}^2<1$.

Since ${\cal D}=0$ for the Lagrangian (\ref{Lagge}), it follows that 
\be
\alpha_{\rm H}=c_{\rm t}^2-1\,,\qquad 
c_{\rm t}^2=F(\phi)\,.
\ee
We also have $q_{\rm t}=M_{\rm pl}^2/8$, $\epsilon_{\S}=0$, 
${\cal W}=2HM_{\rm pl}^2=-{\cal F}$, $\alpha_{\W}=4$, and 
$\epsilon_{\W}=\dot{H}/H^2=-\epsilon_{H}$.
The background equation $L_{,N}+\dot{\cal F}=\rho$, which 
follows from Eqs.~(\ref{back1}) and (\ref{back2}), corresponds to 
$\dot{H}/H^2=-3\Omega_m/2-3\Omega_X$, where 
$\Omega_X=\dot{\phi}^2/(6H^2M_{\rm pl}^2)$ is 
the density parameter of the field kinetic energy.
Then, the scalar sound speed squared (\ref{cs}) reduces to 
\be
c_{\rm s}^2=c_{\rm t}^2
+\frac{3\Omega_m (1-c_{\rm t}^2)
+2\epsilon_{\alpha_{\rm H}}}{6\Omega_X}\,,
\label{csn}
\ee
where $\epsilon_{\alpha_{\rm H}}=2c_{\rm t}\dot{c}_{\rm t}/H$. 
On using Eq.~(\ref{csn}), we can express $q_{\rm s}=\dot{\phi}^2/(2H^2)$ 
without using $\dot{\phi}^2$. 
Then, the quantities $f_1$, $f_2$, and $f_3$ 
in Eqs.~(\ref{f1})-(\ref{f3}) reduce to 
\begin{widetext}
\ba
f_1 &=&
\frac{8M_{\rm pl}^2 c_{\rm t}^2}{c_{\rm s}^2-c_{\rm t}^2} 
\left[ 2c_{\rm s}^2 \epsilon_{\alpha_{\rm H}} 
-2(c_{\rm s}^2-c_{\rm t}^2)(c_{\rm t}^2-1)(c_{\rm t}^2+\epsilon_{\alpha_{\rm H}})
\epsilon_H-3(c_{\rm t}^2-1)c_{\rm t}^2 \Omega_m \right] \nonumber \\
& & -16M_{\rm pl}^2 c_{\rm t}^2(c_{\rm t}^2-1) \left[ (c_{\rm t}^2+\epsilon_{\alpha_{\rm H}})
\epsilon_{\chi}-\epsilon_{\zeta}+(c_{\rm t}^2-1) (\epsilon_H+\epsilon_\chi)
\epsilon_{\zeta} \right]\,,\label{f1d} \\
f_2 &=& \frac{8M_{\rm pl}^2 c_{\rm t}^4}{c_{\rm s}^2-c_{\rm t}^2} 
\left[ 2c_{\rm s}^2 (\epsilon_H-\epsilon_H c_{\rm t}^2+\epsilon_{\alpha_{\rm H}}) 
+c_{\rm t}^2(c_{\rm t}^2-1)(2\epsilon_H-3\Omega_m) \right]
-16M_{\rm pl}^2c_{\rm t}^4 (c_{\rm t}^2-1) (\epsilon_{\chi}-\epsilon_{\zeta})\,,
\label{f2d} \\
f_3 &=& \frac{8M_{\rm pl}^2 c_{\rm t}^2}{c_{\rm s}^2-c_{\rm t}^2} 
\left\{ 2c_{\rm s}^2 [ 1+c_{\rm t}^2(c_{\rm t}^2-2-\epsilon_H
+\epsilon_{\alpha_{\rm H}})+\epsilon_{H}]
-c_{\rm t}^2 (c_{\rm t}^2-1) (2c_{\rm t}^2-2-2\epsilon_H
+3\Omega_m+2\epsilon_{\alpha_{\rm H}}) \right\} \nonumber \\
& &+16 M_{\rm pl}^2 c_{\rm t}^2(c_{\rm t}^2-1)
(c_{\rm t}^2 \epsilon_{\zeta}-\epsilon_{\chi})\,.
\label{f3d}
\ea
\end{widetext}

In the deep matter-dominated epoch characterized by 
$\epsilon_H \simeq 3/2$ and $\Omega_m \simeq 1$, 
we take the quasistatic limits
$\epsilon_{\zeta} \to 0$ and $\epsilon_{\chi} \to 0$ 
in Eqs.~(\ref{f1d})-(\ref{f3d}). 
The validity of this approximation will be checked 
in Sec.~\ref{obsersec}.
Then, the effective gravitational coupling (\ref{quasi1}) and 
the gravitational slip parameter (\ref{quasi2}) reduce, 
respectively, to 
\begin{widetext}
\ba
\hspace{-0.7cm}
\frac{G_{\rm eff}}{G} 
&\simeq& 
1+\frac{1-c_{\rm t}^2}{c_{\rm s}^2}
+\frac{5(1-c_{\rm t}^2)(c_{\rm s}^2-c_{\rm t}^2)}
{c_{\rm s}^2 c_{\rm t}^2(3-3c_{\rm t}^2+2\epsilon_{\alpha_{\rm H}})}
\epsilon_{\alpha_{\rm H}}\,,
\label{Geff2} \\
\hspace{-0.7cm}
\eta 
&\simeq&
1+\frac{5(1-c_{\rm t}^2)(c_{\rm s}^2-c_{\rm t}^2)}
{3c_{\rm t}^2(1+c_{\rm s}^2-c_{\rm t}^2)}
+\frac{25c_{\rm s}^2(c_{\rm s}^2-c_{\rm t}^2)(1-c_{\rm t}^2)}
{3c_{\rm t}^2(c_{\rm t}^2-c_{\rm s}^2-1)
[3c_{\rm t}^6-3c_{\rm t}^4(2+c_{\rm s}^2-\epsilon_{\alpha_{\rm H}})
-3(1+c_{\rm s}^2)c_{\rm t}^2(\epsilon_{\alpha_{\rm H}}-1)
+5c_{\rm s}^2\epsilon_{\alpha_{\rm H}}]}\epsilon_{\alpha_{\rm H}}.
\label{eta2}
\ea
\end{widetext}

When $c_{\rm t}^2$ is constant, i.e., $\epsilon_{\alpha_{\rm H}}=0$, 
these results match with those derived 
in Ref.~\cite{DKT} for scaling solutions realized 
by the potential $V(\phi)=V_1 e^{-\lambda_1 \phi/M_{\rm pl}}$. 
If $c_{\rm t}^2>1$, then $c_{\rm s}^2$ becomes negative 
for $c_{\rm t}^2-1>2\Omega_X/(\Omega_m-2\Omega_X)$. 
Provided that $1<c_{\rm t}^2<1+2\Omega_X/(\Omega_m-2\Omega_X)$, 
we have $G_{\rm eff}<G$ from Eq.~(\ref{Geff2}).
Since $\Omega_X \ll \Omega_m$ during the matter-dominated epoch, 
we require that $c_{\rm t}^2$ is very close to 1 to avoid the 
Laplacian instability of scalar perturbations. 
Then, the deviation of $G_{\rm eff}$ from $G$ is 
restricted to be small. 

If $c_{\tm}^2$ is smaller than 1 without any variation, 
it follows that $G_{\rm eff}>G$. In this case, for $c_{\rm t}^2$ 
away from 1, $c_{\s}^2$ tends to be much larger than 1 
during the matter era. 
Hence $G_{\rm eff}$ is also restricted to be close to $G$. 
In the regime $0<c_{\rm t}^2 \ll 1$, the parameter $\eta$ is
approximately given by 
$\eta \simeq 1+5(1-c_{\rm t}^2)/(3c_{\rm t}^2)$ under
the condition $c_{\rm s}^2 \gg 1$.
This large deviation of $\eta$ from 1 is the distinguished 
observational signature of the model (\ref{Lagge})
with $c_{\rm t}^2$ smaller than 1 \cite{DKT}.
 
If $c_{\rm t}^2$ varies in time, both $G_{\rm eff}$ and $\eta$ 
are subject to change.
If $|\epsilon_{\alpha_{\rm H}}|$ is much smaller than 1, 
we can ignore the contribution of the terms 
$\epsilon_{\alpha_{\rm H}}$ appearing 
in the denominators of Eqs.~(\ref{Geff2}) and (\ref{eta2}). 
Let us consider the case in which $c_{\tm}^2<1$ and  
$c_{\s}^2$ is larger than the order of 1.
Then, Eqs.~(\ref{Geff2}) and (\ref{eta2}) 
approximately reduce to
\ba
\frac{G_{\rm eff}}{G} 
&\simeq& 
1+\frac{1-c_{\rm t}^2}{c_{\rm s}^2}
+\frac{5}{3c_{\rm t}^2}\epsilon_{\alpha_{\rm H}}\,,
\label{Geff3} \\
\eta &\simeq&
1+\frac{5(1-c_{\rm t}^2)}{3c_{\rm t}^2} 
-\frac{25}{9c_{\rm t}^4}\epsilon_{\alpha_{\rm H}}\,.
\label{eta3}
\ea
The decrease (increase) of $c_{\tm}^2$ leads to the negative (positive) 
values of $\epsilon_{\alpha_{\rm H}}=2c_{\tm} \dot{c}_{\tm}/H$.
This means that, even in the subluminal regime ($c_{\tm}^2<1$), 
it is possible to have $G_{\rm eff}<G$ for decreasing $c_{\tm}^2$. 
For $\epsilon_{\alpha_{\rm H}}<0$, the parameter $\eta$ also 
gets increased.

A model with varying $c_{\tm}^2$ can be realized for 
the function
\be
F(\phi)=c_{\tm i}^2\,e^{-2\beta \phi/M_{\rm pl}}\,,
\label{Fphi}
\ee
where $c_{\tm i}$ and $\beta$ are constants. 
For dark energy models in which $\Omega_X$ grows 
in time, Eq.~(\ref{csn}) shows that $|c_{\s}^2|$ tends to 
increase as we go back to the past. This behavior can be 
avoided for scaling dark energy models described
by the field potential \cite{Nunes}
\be
V(\phi)=V_1 e^{-\lambda_1 \phi/M_{\rm pl}}+
V_2 e^{-\lambda_2 \phi/M_{\rm pl}}\,,
\label{potential}
\ee
where $V_1,V_2,\lambda_1,\lambda_2$ are constants 
with $\lambda_1 \gtrsim 10$ and $\lambda_2 \lesssim 1$.
During the matter era, the solution is in the scaling regime 
characterized by $\Omega_X=3/(2\lambda_1^2)$ and 
$\Omega_m=1-3/\lambda_1^2$ \cite{CLW}. 
After the dominance of the second potential on the rhs of Eq.~(\ref{potential}), 
the solution finally approaches an attractor with cosmic acceleration \cite{CST}.

We recall that Eqs.~(\ref{Geff2}) and (\ref{eta2}) are 
valid only during the deep matter era with negligible 
variations of $\zeta$ and $\chi$. 
In order to know the precise evolution of perturbations 
from the matter era to today (especially after the onset of 
cosmic acceleration), we need to resort to numerical simulations.

\section{Observables for a concrete weak gravity model}
\label{obsersec}

In this section we numerically integrate the linear perturbation equations
together with the background equations for the theory given by the Lagrangian 
(\ref{Lagge}). We consider the case in which the functions $F(\phi)$ and 
$V(\phi)$ are given, respectively, by Eqs.~(\ref{Fphi}) and (\ref{potential}). 

\subsection{Background equations and propagation speeds}

In the presence of nonrelativistic matter 
the background quantities 
$x_1 \equiv \dot{\phi}/(\sqrt{6}HM_{\rm pl})$, 
$x_2 \equiv \sqrt{V}/(\sqrt{3}HM_{\rm pl})$, and 
$x_3 \equiv \phi/M_{\rm pl}$ obey the equations of motion
\ba
x_1' &=& \frac32 x_1 \left(x_1^2-x_2^2-1 \right)
+\frac{\sqrt{6}}{2}\lambda x_2^2\,,\\
x_2' &=& \frac32 x_2 \left(x_1^2-x_2^2+1 \right)
-\frac{\sqrt{6}}{2} \lambda x_1 x_2\,,\\
x_3' &=& \sqrt{6}x_1\,,
\ea
where $\lambda \equiv -M_{\rm pl}V_{,\phi}/V$. 
The field equation of state 
$w_{\phi}=[\dot\phi^2/2-V(\phi)]/[\dot\phi^2/2+V(\phi)]$ and 
the matter density parameter 
$\Omega_m=\rho/(3H^2M_{\rm pl}^2)$ 
can be expressed, respectively, as 
\be
w_{\phi}=\frac{x_1^2-x_2^2}{x_1^2+x_2^2}\,,\qquad
\Omega_m=1-x_1^2-x_2^2\,.
\ee

For the potential (\ref{potential}) the scaling matter era 
corresponds to \cite{CLW}
\be
x_1=x_2=\frac{\sqrt{6}}{2\lambda_1}\,,\quad
w_{\phi}=0\,,\quad
\Omega_m=1-\frac{3}{\lambda_1^2}\,,
\label{scaling}
\ee
whereas the late-time scalar-field fixed point 
is characterized by 
\be
x_1=\frac{\lambda_2}{\sqrt{6}}\,,\quad
x_2=\sqrt{1-\frac{\lambda_2^2}{6}}\,,\quad
w_{\phi}=-1+\frac{\lambda_2^2}{3}\,,\quad
\Omega_m=0\,.
\label{scalar}
\ee
{}From the big bang nucleosynthesis bound on $\Omega_m$, 
the slope $\lambda_1$ is constrained 
to be $\lambda_1>9.4$ \cite{Bean}. 
To realize the late-time cosmic acceleration 
we require that $\lambda_2^2<2$, under which 
the scalar-field-dominated fixed point is stable \cite{CST}.

\begin{figure}
\includegraphics[height=3.2in,width=3.3in]{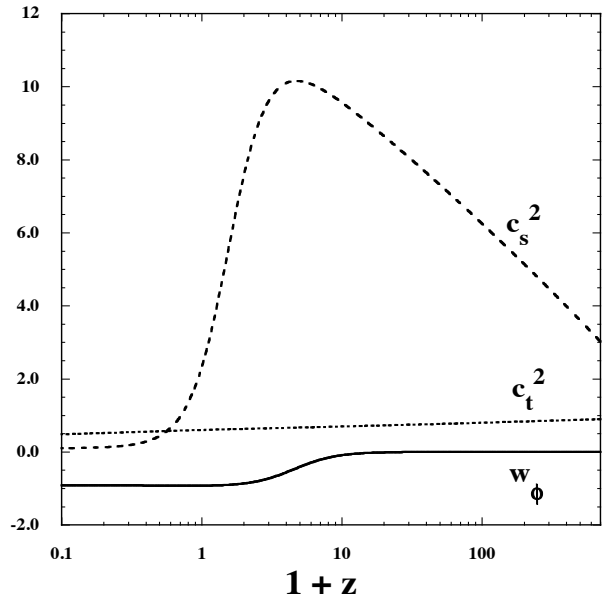}
\caption{\label{fig1}
Evolution of $w_{\phi}, c_{\tm}^2, c_{\s}^2$ versus 
$1+z=1/a$ for $\beta=0.1$, $c_{\tm i}^2=0.9$, $\lambda_1=10$, 
$\lambda_2=0.5$, and $V_2/V_1=10^{-6}$ with the
initial conditions $x_1=\sqrt{6}/(2\lambda_1)$, 
$x_2=\sqrt{6}/(2\lambda_1)$, and $x_3=0$ at $z=735.4$. 
The present epoch (the redshift $z=0$) is identified by 
the condition $\Omega_m=0.3$.}
\end{figure}

In Fig.~\ref{fig1} we plot the evolution of $w_{\phi}$ 
for $\lambda_1=10$,  $\lambda_2=0.5$, $V_2/V_1=10^{-6}$ 
with the initial conditions $x_1=x_2=\sqrt{6}/(2\lambda_1)$ 
and $x_3=0$. 
The field equation of state starts to evolve from 
$w_{\phi}=0$ and then it finally approaches 
$w_{\phi}=-0.917$.
Even if $x_1$ and $x_2$ are initially away from 
these values, the solutions soon approach the scaling 
fixed point (\ref{scaling}) because 
it is a temporal attractor \cite{Nunes}. 
After the dominance of the second potential on the rhs of 
Eq.~(\ref{potential}), the solutions are attracted by the 
fixed point (\ref{scalar}).

Since the scalar field evolves along the potential with  
velocity $\dot{\phi}>0$, the tensor propagation speed 
squared $c_{\tm}^2=c_{\tm i}^2e^{-2\beta \phi/M_{\rm pl}}$ 
decreases from the initial value $c_{\tm i}^2$ for $\beta>0$. 
Provided that $0 \le c_{\tm i}^2 \le 1$, 
$c_{\tm}^2$ always stays in the subluminal 
regime (see Fig.~\ref{fig1} for the case 
$c_{\tm i}^2=0.9$ and $\beta=0.1$).

Let us consider the stability condition associated with 
the scalar propagation speed given by Eq.~(\ref{csn}). 
First of all, the quantity $\epsilon_{\alpha_{\rm H}}$
can be expressed as
\be
\epsilon_{\alpha_{\rm H}}
=-2\sqrt{6} \beta c_{\rm t}^2 x_1\,,
\ee
which is negative for $\beta>0$ and $\dot{\phi}>0$.
Since the deviation of $c_{\tm}^2$ from 1 gives rise to 
negative values of $c_{\s}^2$ for $c_{\tm}^2>1$, 
we focus on the case in which 
$c_{\tm}^2$ is in the subluminal regime.

During the scaling matter era characterized by Eq.~(\ref{scaling}), 
the condition $c_{\s}^2>0$ gives the following bound
\be
\beta<\frac{1}{4\lambda_1} \left[ 3+\left( \lambda_1^2-3 
\right) \left( \frac{1}{c_{\tm}^2}-1 \right) \right]\,.
\label{betacon1}
\ee
When $\lambda_1=10$ and $c_{\tm}^2=0.9$, for example, 
$\beta<0.34$. 

Substituting the values (\ref{scalar}) of the scalar-field-dominated 
fixed point into Eq.~(\ref{csn}), it follows that 
$c_{\s}^2=c_{\tm}^2 (1-4\beta/\lambda_2)$.
If we demand the condition $c_\s^2>0$ for the future attractor, 
we obtain the bound 
\be
\beta<\frac{\lambda_2}{4}\,.
\label{betacon2}
\ee
In the numerical simulation of Fig.~\ref{fig1} the parameter 
$\lambda_2=0.5$ is chosen, so the condition (\ref{betacon2}) translates to 
$\beta<0.125$. We have numerically confirmed that, for $\beta<0.125$,
$c_{\s}^2$ remains positive from the past to the asymptotic future. 
If we do not impose the condition $c_{\rm s}^2>0$ in the future, 
then the bound (\ref{betacon2}) is slightly relaxed, e.g., 
$\beta<0.15$ for $\lambda_2=0.5$.

Figure \ref{fig1} illustrates the evolution of $c_\s^2$ and $c_{\tm}^2$ 
for $\beta=0.1$, $c_{\tm i}^2=0.9$, $\lambda_1=10$, and 
$\lambda_2=0.5$. Since $c_{\tm}^2$ decreases in time, 
this leads to the growth of $c_\s^2$ during the scaling 
matter era. Around the end of the scaling regime the ratio 
$\Omega_m/\Omega_X$ starts to evolve toward 0, 
so $c_{\s}^2$ starts to the decrease. 
Finally, the ratio $c_{\s}^2/c_{\tm}^2$ approaches 
the asymptotic value $1-4\beta/\lambda_2=0.2$.

We have thus clarified the range of $\beta$ in which 
the stability conditions of scalar and tensor perturbations 
are satisfied. Since $q_{\tm}=M_{\rm pl}^2/8$ and $q_{\s}=\dot{\phi}^2/(2H^2)$, 
there are no ghosts in our model.

\subsection{Perturbation equations and initial conditions}

Let us proceed to the discussion of perturbation equations 
and their initial conditions.
Introducing the dimensionless velocity potential 
\be
V_m \equiv Hv\,,
\ee
the perturbation equations of motion can be written as
\ba
\hspace{-0.7cm}
\zeta' &=& \delta N+\frac32 \Omega_m V_m\,,\label{zetaeq}\\
\hspace{-0.7cm}
\chi' &=& \left( \frac{H'}{H}-1 \right) 
\chi-c_{\rm t}^2 \left( \delta N+\zeta \right)\,,\label{tpsi}\\
\hspace{-0.7cm}
\delta' &=& -\frac32 \Omega_m \delta +3(x_1^2-1) 
\delta N+{\cal K}^2 (c_{\rm t}^2 \zeta+V_m)\,,\label{delmde}\\
\hspace{-0.7cm}
V_m' &=&
-\delta N+\frac{H'}{H} 
V_m\,,\label{vmde}\\
\hspace{-0.7cm}
x_1^2 \delta N &=& \frac12 \Omega_m \delta_m
-\frac13 {\cal K}^2 
(c_{\rm t}^2 \zeta+\chi)\,,
\label{delN}
\ea
where ${\cal K} \equiv k/(aH)$ and $H'/H=-(3/2)(1+x_1^2-x_2^2)$.

Taking the ${\cal N}$ derivative of Eq.~(\ref{vmde}) and using 
other equations of motion, we obtain 
\be
V_m''+\alpha_1 V_m'+\alpha_2 V_m=
-{\cal K}^2 \left( \chi-\frac13 \frac{\epsilon_{\alpha_{\rm H}}}
{x_1^2}\zeta \right) \,,
\label{Veq}
\ee
where 
\ba
\alpha_1 &\equiv& 
\frac{\sqrt{6}}{4x_1} \biggl[ 4\lambda (1-\Omega_m-x_1^2)
-\sqrt{6} x_1 (2-\Omega_m-2x_1^2) \biggr]\,,\label{alpha1} 
\nonumber \\ \\
\alpha_2 &\equiv&
(1-c_{\rm t}^2)\frac{\Omega_m}{2x_1^2} {\cal K}^2
\nonumber \\
& &
+\frac{3}{2x_1} \biggl[ 3x_1 \{ 4x_1^4+2(2\Omega_m-3)x_1^2
+\Omega_m (\Omega_m-1) \} 
\nonumber \\
& &
-\sqrt{6}\lambda 
(\Omega_m+4x_1^2)(\Omega_m+x_1^2-1) \biggr]\,.
\label{alpha2}
\ea
The general solution to Eq.~(\ref{Veq}) can be expressed in the 
following form
\be
V_m=V_m^{(h)}+V_m^{(s)}\,,
\ee
where $V_m^{(h)}$ is the homogenous solution satisfying 
the differential equation 
$V_m^{(h)''}+\alpha_1 V_m^{(h)'}+\alpha_2 V_m^{(h)}=0$, 
and $V_m^{(s)}$ is the special solution.

When $c_{\tm}^2<1$ the first term on the rhs of 
Eq.~(\ref{alpha2}) can be much larger than 1 for subhorizon 
perturbations (${\cal K} \gg 1$).
During the scaling matter era characterized by Eq.~(\ref{scaling}), 
the homogenous solution obeys
\be
V_m^{(h)''}+\frac32 V_m^{(h)'}+\left[ c_{\rm eff}^2{\cal K}^2+
\frac{9(\lambda_1^2-3)}{2\lambda_1^2} \right] V_m^{(h)}=0\,,
\ee
where $c_{\rm eff}^2 \equiv (\lambda_1^2-3)(1-c_{\tm}^2)/3$ and 
the quantity ${\cal K}$ evolves as ${\cal K}({\cal N}) \propto e^{{\cal N}/2}$.
If $c_{\tm}^2$ is constant, we obtain the following solution 
in the limit $x \equiv 2c_{\rm eff} {\cal K} \gg 1$ :
\be
V_m^{(h)} \simeq a^{-3/4} \sqrt{\frac{2}{\pi x}} 
\left[ c_1 \cos \left( x-\frac{\pi}{4} \right)+
c_2 \sin \left( x-\frac{\pi}{4} \right) \right]\,,
\label{homoso}
\ee
where $c_1$ and $c_2$ are constants.
Thus, $V_m^{(h)}$ exhibits the oscillation whose frequency is 
related to $c_{\rm eff}$. The amplitude of $V_m^{(h)}$ 
decreases in proportion to $a^{-1}$.

Provided that $c_{\rm t}^2$ is a constant smaller than 1, 
the special solution to Eq.~(\ref{Veq}) is approximately given by 
$V_m^{(s)} \simeq 2x_1^2\,\chi/[\Omega_m (c_{\rm t}^2-1)]$ 
for subhorizon perturbations.
In Ref.~\cite{DKT} it was found that the perturbations $V_m$, $\chi$, 
and $\zeta$ stay nearly constant along the special solution 
during the scaling matter era characterized by Eq.~(\ref{scaling}). 
This means that, as long as the homogeneous solution $V_m^{(h)}$ 
is initially suppressed relative to the special solution $V_m^{(s)}$, 
the latter remains the dominant contribution to $V_m$.
 
For nonconstant $c_{\rm t}^2$, there is an additional term 
on the rhs of Eq.~(\ref{Veq}).
If we neglect the time derivatives of $V_m^{(s)}$ 
for the estimation of the special solution, we obtain  
\ba
V_m^{(s)} &=&
-\frac{{\cal K}^2}{\alpha_2}
\left( \chi-\frac13 \frac{\epsilon_{\alpha_{\rm H}}}{x_1^2}\zeta \right)
 \label{Vss}\\
&\simeq & \frac{2}{c_{\rm t}^2-1} 
\left( \frac{x_1^2}{\Omega_m} \chi-\frac13 
\frac{\epsilon_{\alpha_{\rm H}}}{\Omega_m} \zeta \right)\,,
 \label{Vss2}
\ea
where, in the second line, we have picked up the first term 
on the rhs of Eq.~(\ref{alpha2}).
{}From Eq.~(\ref{Vss2}) we find that the change of $c_{\rm t}^2$
actually gives rise to a variation of $V_m^{(s)}$ 
of the order of $V_m^{(s)'} \simeq 
-\epsilon_{\alpha_{\rm H}}V_m^{(s)}/(c_{\tm}^2-1)$.
This implies that an oscillating mode may arise even 
for initial conditions of $V_m$ that are very close to 
the value (\ref{Vss2}).

\begin{figure}
\includegraphics[height=3.2in,width=3.3in]{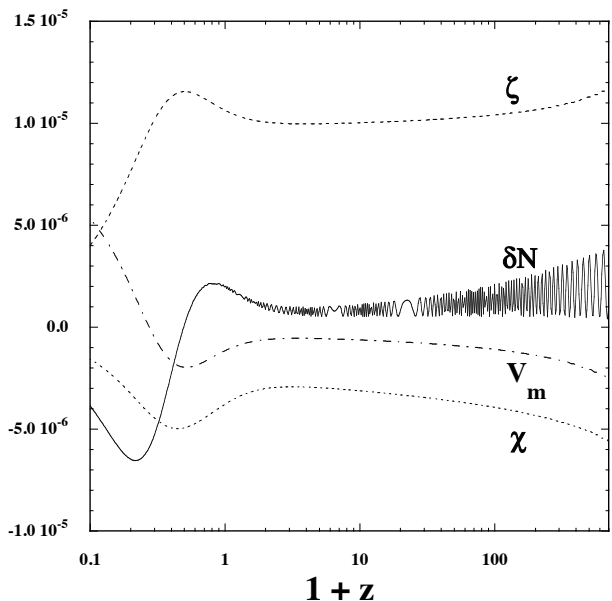}
\caption{\label{fig2}
Evolution of the perturbations $\zeta, \delta N, V_m, \chi$ for 
the same model parameters and background initial conditions 
as those in Fig.~\ref{fig1}.
The initial conditions associated with perturbations are chosen 
to be $\zeta=1.1779 \times 10^{-5}$, $\chi=-5.6540 \times 10^{-6}$, 
$\delta=2.1253 \times 10^{-3}$,  $V_m=-2.6174 \times 10^{-6}$, 
and ${\cal K}=25$ at redshift $z=735.4$. }
\end{figure}

We recall that the parameter $\beta$ is bounded from above, see  
Eqs.~(\ref{betacon1}) and (\ref{betacon2}). 
In particular the slope $\lambda_2$ needs to be smaller than the order of 1
in order to be consistent with observations of cosmic acceleration, 
so the condition (\ref{betacon2}) gives the bound $\beta <O(0.1)$. 
During the scaling matter era the parameter $\epsilon_{\alpha_{\rm H}}$ 
is given by $\epsilon_{\alpha_{\rm H}}=-6c_{\rm t}^2 \beta/\lambda_1$, 
so it is smaller than the order of 0.1.  
Then the terms $\alpha_1 V_m^{(s)'}$ and $V_m^{(s)''}$ 
in Eq.~(\ref{Veq}) are not as large as $\alpha_2 V_m^{(s)}$ 
for $c_{\tm}^2$ not very close to 1.

Numerically we solve the perturbation equations of motion 
for the initial conditions $V_m=V_m^{(s)}$, $V_m'=0$, and $\chi'=0$, 
where $V_m^{(s)}$ is given by Eq.~(\ref{Vss}). 
For given $\delta_m$, the initial values of $\delta N$, $\zeta$ and $\chi$ are known 
accordingly from Eqs.~(\ref{tpsi}), (\ref{vmde}), and (\ref{delN}). 
The initial condition of $\delta_m$ is chosen such that its value today
is equivalent to $\sigma_8(0)=0.82$, where $\sigma_8(0)$ is the rms amplitude 
of overdensity at the comoving $8h^{-1}$ Mpc scale 
($h$ is the normalized Hubble parameter 
$H_0 = 100\,h$ km ${\rm sec}^{-1}{\rm Mpc}^{-1}$). 

In Fig.~\ref{fig2} we plot the evolution of $\zeta$, $\chi$, $V_m$, and 
$\delta N$ for the same model parameters 
as those given in Fig.~\ref{fig1}. 
Unlike the constant $c_{\tm}^2$ model \cite{DKT}, 
the perturbations $V_m$, $\zeta$, and $\chi$ do not stay constant even during the 
scaling matter era. This variation is induced by the change of $c_{\tm}^2$ 
appearing in the special solution (\ref{Vss2}).

In Fig.~\ref{fig2} the perturbation $\delta N$ exhibits
damped oscillations with a non-negligible initial amplitude.
This comes from the fact that $\delta N$ 
is related to the derivative $V_m'$, as 
$\delta N=-V_m'+(H'/H)V_m$. 
We recall that, even by choosing an initial value of $V_m$ that 
is identical to $V_m^{(s)}$ in Eq.~(\ref{Vss}), 
the contribution to $V_m$ from the homogenous 
solution $V_m^{(h)}$ arises due to the variation of $c_{\tm}^2$. 
Taking the ${\cal N}$ derivative of Eq.~(\ref{homoso}), 
it follows that the amplitude of $V_m^{(h)'}$
is $x/2$ times as large as that of $V_m^{(h)}$.
Since we are considering the case $x \gg 1$,
the oscillating mode particularly 
manifests itself in the perturbation $\delta N$ 
through the derivative $V_m^{(h)'}$. 

For initial conditions of $V_m$ away from the special 
solution $V_m^{(s)}$, the perturbations exhibit damped 
oscillations with larger amplitudes than those plotted in Fig.~\ref{fig2}. 
This situation is similar to what happens for the 
constant $c_{\tm}^2$ model \cite{DKT}.
Since it is likely that such large oscillations can be 
severely constrained from CMB observations, 
we focus on the case in which $V_m$ is initially 
close to $V_m^{(s)}$ in the following discussion.

 \subsection{Observables}
 \label{observasec}
\begin{figure}
\includegraphics[height=3.2in,width=3.3in]{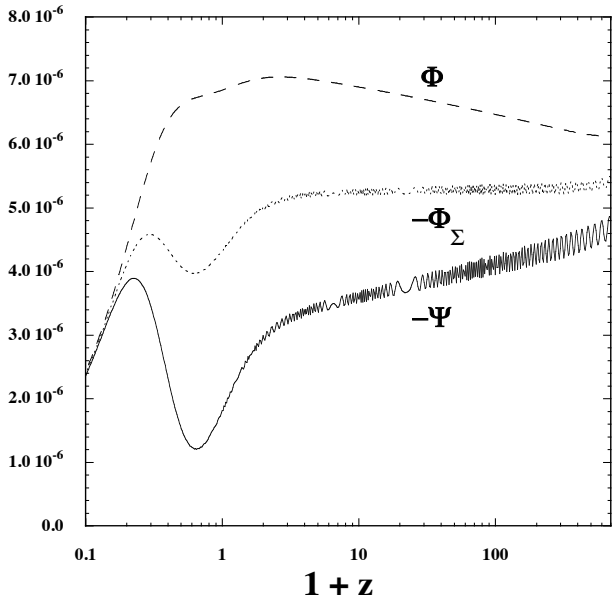}
\caption{\label{fig3}
Evolution of the gravitational potentials $\Phi$, $-\Psi$, $-\Phi_{\Sigma}$ 
for the same model parameters and initial conditions as those given 
in the captions of Figs.\,\ref{fig1} and \ref{fig2}.}
\end{figure}
\begin{figure}
\includegraphics[height=3.2in,width=3.1in]{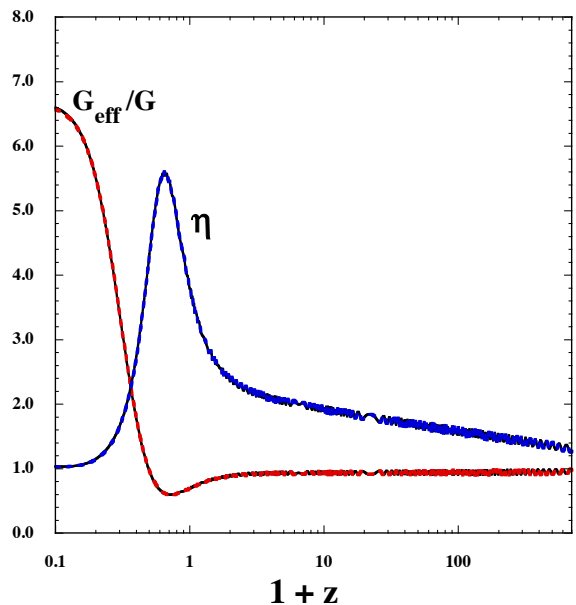}
\caption{\label{fig4}
Evolution of $G_{\rm eff}/G$ and $\eta$ (black solid lines)
for the same model parameters and initial conditions 
as those given in the captions of Figs.~\ref{fig1} and \ref{fig2}.
The dashed red and blue lines correspond to the evolution 
of $G_{\rm eff}/G$ and $\eta$, respectively, 
derived under the subhorizon approximation 
involving $\epsilon_{\zeta}$ and $\epsilon_{\chi}$, 
i.e., Eqs.~(\ref{quasi1}) and (\ref{quasi2}). }
\end{figure}

We study the evolution of observables associated with RSD, 
weak lensing, and CMB.
In Fig.~\ref{fig3} we plot the gauge-invariant 
gravitational potentials $-\Psi$ and $\Phi$ 
as well as $-\Phi_{\Sigma}$ versus $1+z$ 
for the same model parameters 
and initial conditions as those used in 
Figs.~\ref{fig1} and \ref{fig2}.
Since $\Psi=\delta N+\chi'-(H'/H)\chi$, the oscillation of 
$\delta N$ seen in Fig.~\ref{fig2} leads to that of 
$-\Psi$. In Fig.~\ref{fig3} the oscillating amplitude 
of $-\Psi$ is not so large due to the choice of the initial 
condition $|V_m^{(h)}| \ll |V_m^{(s)}|$. 
With the decrease of $V_m^{(h)'}$,  
the oscillations of $-\Psi$ damp away. 
The gravitational potential $-\Psi$ exhibits 
an overall decrease from the matter era to today, so this case 
corresponds to weak gravity.

Figure \ref{fig4} shows the evolution of $G_{\rm eff}/G$ 
computed from the definition of $G_{\rm eff}$
given in Eq.~(\ref{Geffdef}). 
{}From the scaling matter era to today we have $G_{\rm eff}/G<1$,  
while satisfying the no-ghost and stability conditions of tensor and 
scalar perturbations (see Fig.~\ref{fig1}). 
The realization of weak gravity comes from the fact that 
the last term on the rhs of Eq.~(\ref{Geff2}) is negative due to 
the decrease of $c_{\tm}^2$. 
For constant $c_{\tm}^2$ smaller than 1, $G_{\rm eff}$ 
is larger than $G$ for subhorizon perturbations.
 
In Fig.~\ref{fig4} we also plot $G_{\rm eff}/G$ given in Eq.~(\ref{quasi1})  
derived under the subhorizon approximation. 
We have computed the terms $\epsilon_{\zeta}$ and $\epsilon_{\chi}$ 
by solving the full perturbation equations and then substituted them 
into Eqs.~(\ref{f1}) and (\ref{f2}). 
The numerical simulation of Fig.~\ref{fig4} shows that 
the approximate formula (\ref{quasi1})
can reproduce the numerical results with high accuracy 
for the modes deep inside the Hubble radius.
Even by setting $\epsilon_{\zeta}=0=\epsilon_{\chi}$ 
in Eqs.~(\ref{f1}) and (\ref{f2}) we find that 
the analytic formula of $G_{\rm eff}/G$ is a good 
approximation during the scaling matter era, but 
it starts to deviate from the full numerical integration 
around the end of the matter era. 
In GLPV theories the quasistatic approximation ignoring 
the variation of $\epsilon_{\zeta}$ and $\epsilon_{\chi}$ is not trustable 
at the late cosmological epoch.

In the regime where $c_{\s}^2$ is larger than the order of 1, 
the gravitational slip parameter is given by Eq.~(\ref{eta3}) 
under the subhorizon approximation in the deep matter era.
Since the second and third terms on the rhs of Eq.~(\ref{eta3}) 
are positive for $c_{\tm}^2<1$ and $\epsilon_{\alpha_{\rm H}}<0$, 
$\eta$ is larger than 1. 
In Fig.~\ref{fig4} we find that $\eta$
grows as the decrease of $c_{\tm}^2$ from the matter era 
to today and it starts to decrease in the future. 
This deviation of $\eta$ from 1 is one of the distinguishing
features of our model.
As we see in Fig.~\ref{fig4}, the subhorizon approximation 
based on Eq.~(\ref{quasi2}) with inclusion of $\epsilon_{\zeta}$ 
and $\epsilon_{\chi}$ is a trustable prescription for the 
evolution of $\eta$. 

Since $\eta$ is larger than 1, the two gravitational potentials 
$\Phi$ and $-\Psi$ are different from each other. 
In Fig.~\ref{fig3} we find that $\Phi$ grows during the 
matter era, while $-\Psi$ decreases.  
We recall that the weak lensing gravitational potential 
$\Phi_{\Sigma}$ obeys Eq.~(\ref{PhiSigma}) with 
$\Sigma$ given by Eq.~(\ref{Sigma}). 
Since $G_{\rm eff}<G$ and $\eta>1$ during the matter era 
in the numerical simulation of Figs.~\ref{fig3} and \ref{fig4}, 
the small value of $G_{\rm eff}$ appearing in $\Sigma$ is 
compensated by the large value of $\eta$. 
Hence $-\Phi_{\Sigma}$ does not vary much relative to 
$\Phi$ and $-\Psi$ (see Fig.~\ref{fig3}).

\begin{figure}
\includegraphics[height=3.1in,width=3.3in]{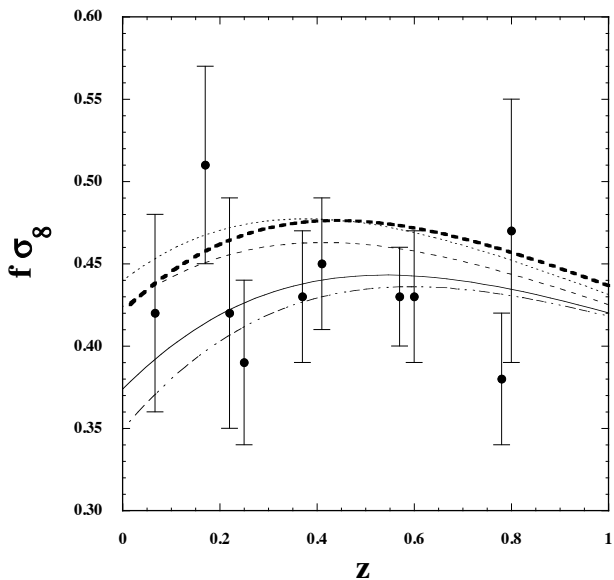}
\caption{\label{fig5}
Evolution of $f\sigma_8$ in the redshift range $0 \le z \le 1$ for 
$c_{{\rm t}i}^2=0.9$, $\lambda_1=10$, 
$\lambda_2=0.5$, $V_2/V_1=10^{-6}$, $\sigma_8(0)=0.82$, 
and $\beta=0, 0.05, 0.1, 0.12$ (from top to bottom) with the 
initial conditions $x_1=\sqrt{6}/(2\lambda_1)$, 
$x_2=\sqrt{6}/(2\lambda_1)$, $x_3=0$, and ${\cal K}=25$ 
at $z=735.4$. The initial conditions of perturbations are 
chosen to satisfy $|V_m^{(h)}| \ll |V_m^{(s)}|$. 
The bold dotted curve corresponds to the evolution of 
$f\sigma_8$ in the $\Lambda$CDM model.
The data points with error bars come from the recent observations of 
$f(z)\sigma_8 (z)$ by 6dFGRS, 2dFGRS, WiggleZ, SDSSLRG, 
BOSSCMASS, and VIPERS surveys.}
\end{figure}

The growth rate of matter density perturbations $\delta_m$ can be measured 
by peculiar velocities of galaxies in RSD surveys \cite{Kaiser,Tegmark}. 
Usually, this is quantified by the data of $f(z) \sigma_8(z)$ at redshift $z$, 
where $f=\dot{\delta}_m/(H\delta_m)$.
In Fig.~\ref{fig5} we plot the ten data points of $f(z) \sigma_8(z)$ 
with error bars derived from the measurements of 2dFGRS \cite{2dF}, 6dFGRS \cite{6dF}, 
WiggleZ \cite{Wiggle}, SDSSLRG \cite{SDSS}, 
BOSSCMASS \cite{BOSS}, and VIPERS \cite{Torre}. 
The latest Planck measurement of the CMB power spectra 
provided the bound $\sigma_8(0)=0.829 \pm 0.014$ at the 68 \% 
confidence level \cite{Planck2}. 

In Fig.~\ref{fig5} we also show the theoretical curves for $\beta=0, 0.05, 0.1, 0.12$, 
$c_{{\rm t}i}^2=0.9$, and 
$\sigma_8 (0)=0.82$ with the initial conditions of perturbations 
satisfying $|V_m^{(h)}| \ll |V_m^{(s)}|$. 
When $\beta=0$ we have $\epsilon_{\alpha_{\rm H}}=0$ and 
$G_{\rm eff}/G \simeq 1+(1-c_{\tm}^2)/c_{\s}^2$ 
in the deep matter era, but $G_{\rm eff}$ is close to $G$ 
due to the condition $c_{\s}^2 \gg 1$ \cite{DKT}. 
In Fig.~\ref{fig5} the theoretical curves for $\beta=0$ and 
the $\Lambda$CDM model do not fit 
the recent RSD data very well.

When $\beta>0$ the effective gravitational coupling $G_{\rm eff}$ is 
smaller than $G$ from the matter era to today, so 
the growth rate of $\delta_m$ gets smaller relative to the case $\beta=0$. 
For larger $\beta$, the theoretical curves shift toward smaller 
values of $f\sigma_8$. {}From Fig.~\ref{fig5} we find that the varying 
$c_{\tm}^2$ models with $\beta>0.05$ show a better agreement with 
the RSD data compared to the constant $c_{\tm}^2$ model. 
We recall that $\beta$ is constrained as $\beta<\lambda_2/4$ 
to avoid $c_{\s}^2<0$ for the future attractor, i.e., 
$\beta<0.125$ in the numerical simulation of Fig.~\ref{fig5}. 
In the regime $0.05<\beta<0.125$, the models are compatible with 
the RSD measurements for the values of 
$\sigma_8 (0)$ constrained by Planck.

\section{Conclusions}
\label{consec}

In this paper we have studied the possibility of realizing 
a gravitational interaction weaker than that in GR on scales 
relevant to large-scale structures and 
weak lensing. We have employed the approach of the effective field theory 
of modified gravity encompassing both Horndeski and GLPV theories as 
specific cases. The important quantities associated with the no-ghost and stability 
conditions of tensor and scalar perturbations are given 
by Eqs.~(\ref{qtctdef}), (\ref{scaghost}), (\ref{cs2}).
All of the quantities $q_{\rm t}$, $c_{\rm t}^2$, 
$q_{\rm s}$, and $c_{\rm s}^2$ are required to be positive for 
theoretical consistency.

Since our interest is the evolution of perturbations for modes deep 
inside the Hubble radius ($k/a \gg H$), we have exploited 
the subhorizon approximation under which the dominant contributions 
to the perturbation equations are those involving $k^2/a^2$ and 
the matter density perturbation $\delta \rho$. 
In Sec.~\ref{subsec} we have derived the general expressions of the 
effective gravitational coupling $G_{\rm eff}$, gravitational slip parameter 
$\eta$, and weak lensing parameter $\Sigma$ by using the quantities 
$q_{\tm}$, $c_{\tm}^2$, $q_{\s}$, and $c_{\s}^2$.
In GLPV theories the time derivatives $\dot{\zeta}$ and $\dot{\chi}$ 
do not vanish even after employing the subhorizon approximation, so 
we need to know the numerical values of $\epsilon_{\zeta}$ and 
$\epsilon_{\chi}$ in Eqs.~(\ref{f1})-(\ref{f3}) 
for the computations of $G_{\rm eff}$, $\eta$, and $\Sigma$ 
(unless $|\epsilon_{\zeta}|$ and $|\epsilon_{\chi}|$ 
are much smaller than unity).

In Horndeski theories the analytic expressions of $G_{\rm eff}$, $\eta$, 
and $\Sigma$ are of the simple forms (\ref{GeffHo})-(\ref{SigmaHo}).
In this case the necessary condition for the realization of weak gravity 
is given by Eq.~(\ref{ctcon}), but this is not a sufficient condition 
due to the presence of additional scalar-matter interactions.
The scalar-matter coupling, which always enhances $G_{\rm eff}$, 
should not be so large as to give rise to values 
of $G_{\rm eff}$ larger than $G$. 

In GLPV theories the two conditions (\ref{ABcon}) are absent, so 
$c_{\tm}^2$ can deviate from 1 even in the deep matter era.
We have presented a simple time-varying $c_{\tm}^2$ model 
described by the Lagrangian (\ref{Lagge}) with the function (\ref{Fphi}). 
In this model the scalar propagation speed squared $c_{\s}^2$, which 
is given by Eq.~(\ref{csn}), increases to be much larger than 1 
for dark energy models with decreasing $\Omega_X$ toward the past. 
For scaling models described by the field potential (\ref{potential})
the ratio $\Omega_m/\Omega_X$ stays constant during the matter era, 
so the unbound growth of $c_{\s}^2$ in the past can be avoided.

The analytic expression of $G_{\rm eff}/G$ given in Eq.~(\ref{Geff3}), 
which is valid in the deep matter era with $c_{\s}^2 \gg 1$ and 
$|\epsilon_{\zeta}| \ll 1, |\epsilon_{\chi}| \ll 1$, shows that 
it is possible to realize weak gravity ($G_{\rm eff}<G$) 
for the decreasing $c_{\tm}^2$ model ($\epsilon_{\alpha_{\rm H}}<0$) 
in the regime $0<c_{\tm}^2<1$. 
In this case, the parameter $\eta$ is 
larger than 1 from Eq.~(\ref{eta3}).

In Sec.~\ref{obsersec} we have numerically solved the full perturbation 
equations of motion for the model (\ref{Lagge}) with the functions (\ref{Fphi}) and 
(\ref{potential}). The solution to the normalized velocity potential 
$V_m=Hv$ can be written by the sum of the homogenous solution 
$V_m^{(h)}$ and the special solution $V_m^{(s)}$.
Provided that $V_m^{(h)}$ is initially much smaller than $V_m^{(s)}$, 
the amplitudes of oscillating modes of perturbations are suppressed apart from 
$\delta N$ and $\Psi$ that are related to the time derivative of $V_m^{(h)}$.
The numerical simulations of Figs.~\ref{fig3} and \ref{fig4} show the 
realization of weak gravity from the matter era to today with $\eta$
larger than 1.

In Fig.~\ref{fig5} we have also computed the evolution of $f\sigma_8$ 
in the redshift range $0 \le z \le 1$ for several different values of $\beta$.
For larger $\beta$ the theoretical values of $f\sigma_8$ get smaller, so 
that these cases exhibit better compatibility with the recent RSD data relative 
to the case $\beta=0$. It remains to be seen whether or not 
the future RSD data combined with other observational probes 
favor the lower growth rate of matter perturbations than 
that in the $\Lambda$CDM model.

There are several issues we have not addressed in this paper.
First, under the so-called disformal transformation \cite{Beken,Garcia,Sasaki}, 
the Lagrangian (\ref{Lagge}) can be transformed to the one in the Einstein frame 
in which the tensor propagation speed 
is 1 \cite{Cremi,DeFelice,Tsuji14}. 
Since a nontrivial kinetic coupling with the scalar field and matter arises in the 
Einstein frame \cite{Mota,Hami2,Bettoni}, the role of such an interaction should be understood in view of a coupled dark energy and dark matter scenario \cite{inter}.

Moreover, it will be of interest to study the screening 
mechanism of the fifth force \cite{KWY} in local regions of the Universe
for the model (\ref{Lagge}).
In some modified gravity models like the quartic Galileon, 
it was shown that the screening mechanism can give rise to 
a $G_{\rm eff}$ smaller than $G$ in the nonlinear regime 
of matter perturbations \cite{quartic}.
It will be of interest to see whether such properties persist 
in more general modified gravity models in the framework 
of GLPV theories.

\section*{ACKNOWLEDGEMENTS}
The authors thanks Antonio De Felice, Kazuya Koyama, and 
Federico Piazza for useful discussions. 
The author is supported by the Grant-in-Aid for Scientific 
Research Fund of the JSPS No.\,24540286 and the Fund for Scientific Research 
on Innovative Areas (JSPS No.\,21111006). 


\end{document}